\documentclass[journal]{vgtc}                     

\onlineid{0}

\vgtccategory{Research}

\vgtcpapertype{please specify}

\title{3D Gaussian Modeling and Ray Marching of OpenVDB datasets for Scientific Visualization}

\author{%
  \authororcid{Isha Sharma}{0000-0001-7978-3906},
  \authororcid{Dieter Schmalstieg} {0000-0003-2813-2235}
}

\abstract{%
  3D Gaussians are currently being heavily investigated for their scene modeling and compression abilities. In 3D volumes, their use is being explored for representing dense volumes as sparsely as possible. However, most of these methods begin with a memory inefficient data format. Specially in Scientific Visualization(SciVis), where most popular formats are dense-grid data structures that store every grid cell, irrespective of its contribution. OpenVDB library\cite{OpenVDB} and data format were introduced for representing sparse volumetric data specifically for visual effects use cases such as clouds, fire, fluids etc. It avoid storing empty cells by masking them during storage. It presents an opportunity for use in SciVis, specifically as a modeling framework for conversion to 3D Gaussian particles for further compression and for a unified modeling approach for different scientific volume types. This compression head-start is non-trivial and this paper would like to present this with a rendering algorithm based on line integration implemented in OptiX8.1\cite{NVIDIAOptiX8Documentation} for calculating 3D Gaussians contribution along a ray for optical-depth accumulation. For comparing the rendering results of our ray marching Gaussians renderer, we also implement a SciVis style primary-ray only NanoVDB HDDA based ray marcher for OpenVDB voxel grids. Finally, this paper also explores application of this Gaussian model to formats of volumes other than regular grids, such as AMR volumes and point clouds, using internal representation of OpenVDB grid class types for data hierarchy and subdivision structure. 
}


\teaser{
	\centering
	\includegraphics[width=\linewidth]{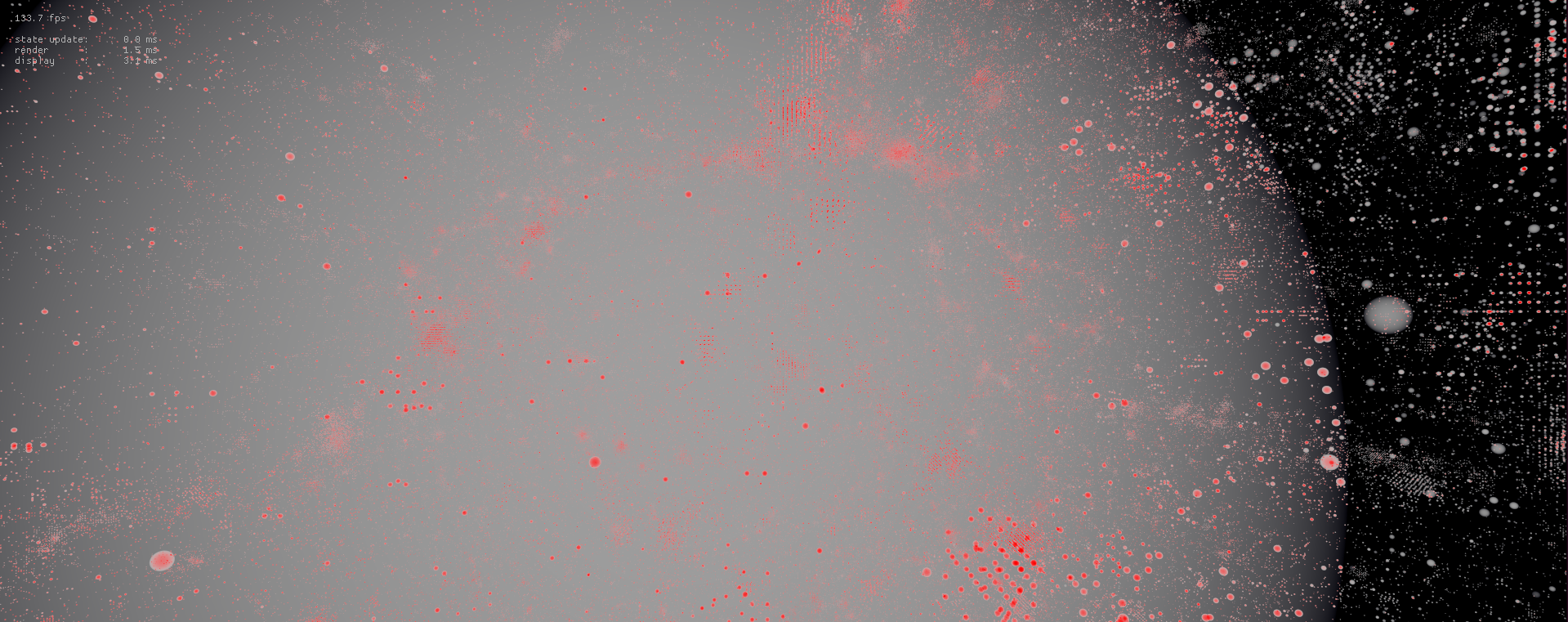}\\[-1pt]
	\includegraphics[width=\linewidth]{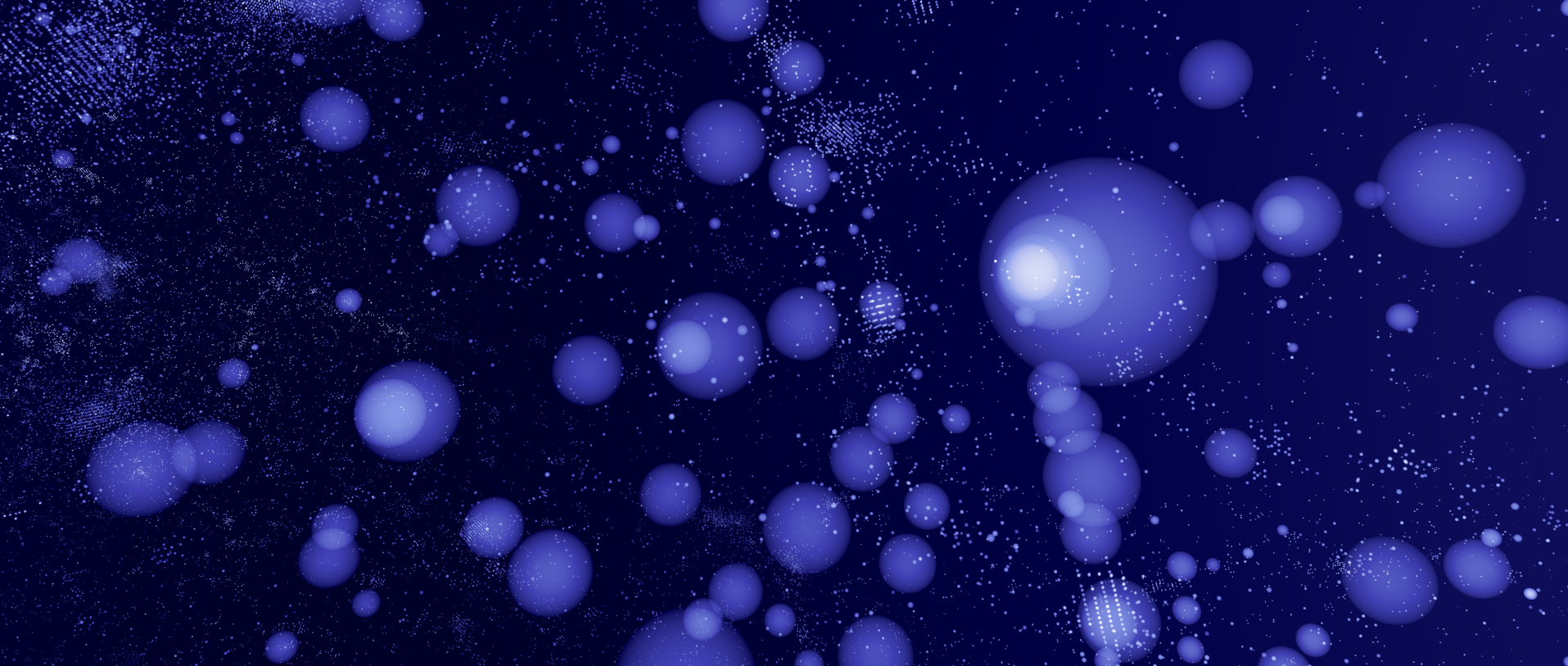}
	\caption{%
    Rendering of the \textit{darksky\_1M.xyz} particle dataset~\cite{NVIDIA2022DarkSky1M}, converted into an OpenVDB PointDataGrid, where each particle is represented as a 3D volumetric Gaussian. Top: transfer function mapping from white to red on a black background, with a particle placed very close and in direct line-of-sight of the camera. Bottom: transfer function mapping from blue to white producing a soft blue periphery of a Gaussian located near the camera at the right side.
	}
	\label{fig:teaser}
}

\renewcommand{\manuscriptnotetxt}{}

\nocopyrightspace

\graphicspath{{figs/}{figures/}{pictures/}{images/}{./}} 

\usepackage{tabu}                      
\usepackage{booktabs}                  
\usepackage{lipsum}                    
\usepackage{mwe}                       

\usepackage{amsmath}
\usepackage{algorithm}
\usepackage[noend]{algpseudocode}

\usepackage{fancyhdr}
\begin{document}



\maketitle

\section{Introduction} 
Ray marching of volumetric data is one of the standard ways of rendering grid-based volumes, also known as Direct Volume Rendering\cite{DirectVolumeRendering2007}. Grids are a collection of ``voxels'', which are cuboidal cell equivalent of a pixel for a volume. Often times they are regular and structured, with equal voxels. But there are also grids with voxels of different resolution, shapes and sizes. Conversion of a ``voxel-space'' to a set of 3D Gaussians, that exist as particles with 3D coordinates, a mean value and spread factor(s) in world space, to provide the same accumulated optical depth has been attempted multiple times(please see related works section). However, the inherent structural dissimilarity and differing amount of region, a voxel and a Gaussian ideally encode, it's a challenge to maintain close fidelity between the two representations. Hence, ray marching of 3D Gaussians although conceptually similar to voxels, is mathematically quite different and in this paper we investigate one such technique applied to OpenVDB\cite{OpenVDB} volumes converted 3D Gaussians, that are then evaluated using line integration\cite{inproceedings} along a ray in OptiX8.1\cite{NVIDIAOptiX8Documentation} for rendering. 

 Most of the recent real-time 3D Gaussian particle ``surface'' representation techniques like in 3DGS\cite{kerbl3Dgaussians} have focused on storing a 'learned' color information or a view dependent appearance value using spherical harmonics which enable advanced lighting effects like specular highlights. The values are learned through a training component which is tightly coupled with the differentiable rasterization that enables this optimization of per-Gaussian appearance and geometry by backpropagating image reconstruction loss through the rendering process. Once the model is trained all the lighting and color information is baked and cannot be changed at run-time or without going through the training loop again, making it a static representation. Optimized Gaussians provide high quality rendering using rasterization, however have an initial cost of training. Hence, any lighting or environment changes require recalculating the Gaussian properties by going through the training loop which has a substantial time overhead to it.   

 Visualization of a volumetric dataset requires application of a transfer function that converts the accumulated optical depth post-traversal to distinguishable colors. The 3DGS techniques for storing final color information in the Gaussian properties, have hence inspired baking in transfer function color information for volumetric datasets in SciVis or related volumetric visualization fields such as Medical Visualization\cite{DBLP:conf/vmv/NiedermayrNPEW24}. Data points with single or multiple: scalar, vector or tensor quantities are encoded using a transfer function to a color and/or opacity value prior to training of the Gaussian Model. Once the color information is baked into the model, changing or adjusting the view dependent features require re-training.
 Some techniques have trained on multiple transfer functions for accommodating larger set of transfer function(trained and untrained)\cite{Dyken2025VolumeEncodingGaussians}, theoretically it doesn't make the technique transfer function ``agnostic'' but more flexible. 
 
In contrast to optimization-based methods, our approach\footnote{This work is submitted to the IEEE for possible publication. Copyright may be transferred without notice, this version may no longer be accessible later.} constructs the Gaussian representation directly from the original volumetric data, without applying a transfer function during initialization. Rather than relying on a training loop to adjust Gaussian properties, we initialize all attributes such as position, covariance, and opacity, based on the raw voxel values and spatial distribution provided by OpenVDB. This avoids the computational cost associated with differentiable rasterization and gradient-based optimization typically required to fit Gaussians from image supervision, also circumventing any need for structure-from-motion preprocessing(e.g., COLMAP) or multi-view image inputs. Visualization is done by accumulating optical depth via ray marching through the Gaussian model and applying the transfer function in the rendering loop to produce the final color, similar to how it is done in a standard volume ray marching renderer. 

OpenVDB is a hierarchical data storage format for sparse volumes and like most other spatial subdivision formats, it employs a tree like structure with variable number of nodes at each level, albeit all nodes of a level have same resolution. It supports sparse nodes by masking inactive parts of the volume during storage using bit-masks. Only the ``active'' voxels or even complete nodes can hence be accessed during traversal and used for Gaussian initialization. This eradicates a large portion of otherwise non-contributing data values which are hallmark of dense grid storage formats. 

Due to the particle and multi-resolute nature of 3D Gaussians, we can effectively convert multiple different volumetric/point-based OpenVDB grid representations using the same strategy and prove the versatility of this data representation model.  

To summarize, the contribution of this paper are the following: 

\begin{itemize}
\item{Ray marching renderer using line integration\cite{inproceedings} of a ray through 3D Gaussian model of a volume implemented in OptiX}
\item{Transfer function agnostic 3D Gaussian model derivation from OpenVDB Tree data structure}
\item{Showcasing this Gaussian model's ability to render AMR volumes via OpenVDB multiple resolution-varying grids and particles via OpenVDB PointDataGrid}
\end{itemize}

\section{Related Work}
3D scene re-modeling and fitting with Gaussians have been studied and applied for Visualization multiple times in the past. Jang et al.'s\cite{jang2006enhancing} work develops a volumetric approximation and visualization system using ellipsoidal Gaussian basis functions to compactly represent scattered volumetric data. In their findings, ellipsoidal Gaussians achieve better compression compared to spherical Gaussians and more accurate representation of structures that are non-spherical. They also remain efficient for GPU-based rendering. Similarly, Vucini et al.'s\cite{Vucini2009PhD} work on 3D reconstruction and visualization from non-uniform scalar and vector data, explored benefits of using radial basis functions and Gaussian primitives to address the challenges in adapting irregular sampling to GPU-friendly cartesian or regular grids. Similar to our motivation, Juba et al.'s\cite{JubaVarshney2007GaussianRBFVolume} work introduced a multi-resolution implicit representation of scalar volumetric data over an octree, using anisotropic Gaussian radial basis functions (RBFs) with an efficient MLE-based encoding algorithm. In addition, they also experimented with level-of-detail control in their GPU-accelerated ray-casting method that enabled direct rendering, hence achieving interactive performance even on very large datasets. This 2002 study by Lazzaro et al. \cite{Lazzaro2002RBFInterpolation} incorporates radial basis functions (RBFs) for multivariate interpolation of very large scattered datasets, into a modified Shepard’s scheme. All the above listed works are atleast a decade old and GPU capabilities have exploded in recent past making Gaussian processing and rendering much faster.  Among some of the recent works include Volume Encoding Gaussians\cite{Dyken2025VolumeEncodingGaussians} that tailor 3DGS work towards Scientific Visualization.

Most OpenVDB-based implementations have concentrated on efficiently rendering photorealistic volumetric content, with notable integration into production tools such as Blender, as well as proprietary systems such as Houdini and platforms like Omniverse\cite{NvidiaOmniverse} developed by NVIDIA. However, several studies have also explored the use of OpenVDB in Scientific Visualization(SciVis), medical imaging, and large-scale simulation rendering. Mayer et al.~\cite{mayer2021visualization} present a method for visualizing human-scale blood flow simulations using Intel OSPRay Studio on the SuperMUC-NG supercomputer. By mapping simulation data to memory-efficient VDB volumes, they enabled interactive visualization without requiring extensive preprocessing. Vizzo et al.~\cite{vizzo2022vdbfusion} introduce \textit{VDBFusion}, a system that integrates range sensor data into truncated signed distance fields(TSDFs) using OpenVDB. Their implementation achieves real-time LiDAR data processing at 20 frames per second on a single-core CPU, demonstrating OpenVDB’s efficiency for robotics applications. Bailey et al.~\cite{bailey2015distributing} propose a framework for integrating OpenVDB with OpenMPI to distribute liquid simulations across multiple processors. This solution targets efficient simulation of complex fluid dynamics in visual effects workflows.  Walker et al.~\cite{walker2022nanomap} present NanoMap, a GPU-accelerated mapping and simulation library that combines OpenVDB and CUDA to process dense point clouds for robotic agents. The system significantly improves real-time occupancy mapping and simulation, particularly on resource-constrained platforms. In the context of science communication, Borkiewicz et al.~\cite{borkiewicz2017communicating} highlight the importance of visualization in effectively conveying scientific ideas, especially in an era challenged by misinformation. They emphasize the need to balance scientific rigor with visually engaging narratives to improve public understanding. In terms of Level-of-Detail(LOD) techniques using Gaussian primitives, Seo et al.~\cite{seo2024flod} introduced FLoD, a flexible LOD scheme for 3D Gaussian Splatting. While their approach targets splatting-based models, our work presents a novel alternative: generating LOD-controllable Gaussian approximations directly from OpenVDB data. This expands the applicability of OpenVDB in rendering pipelines beyond traditional surfaces and volumes.

Recent work on Gaussian ray tracing~\cite{MoenneLoccoz2024} introduces efficient acceleration structures for ray tracing particle scenes, by first enclosing Gaussians within bounding geometries before placing them into axis-aligned bounding boxes (AABBs) for OptiX BVH construction. The method is primarily targeted at photorealistic 3D scene reconstruction, deriving and optimizing Gaussians directly from image data within the same end-to-end pipeline as 3DGS~\cite{kerbl20233dgaussiansplattingrealtime}. While the focus differs from our purely volumetric datasets rendering, the use of bounding geometries for Gaussians demonstrates transferable strategies for improving both rendering performance and quality in our method. Along similar lines of dealing with volumetric Gaussians in a consistent way for volume-rendering, but during rasterization Talegaonkar et al.\cite{talegaonkar2025vol3dgs} proposes enhancing the physical fidelity of Gaussian splatting methods (e.g., 3DGS) by analytically integrating 3D Gaussians within a rasterization framework instead of relying on screen-space splatting approximations. They derive transmittance in a closed-form that lead to physically accurate alpha values, which can be then incorporated in a 3DGS pipeline.

\section{Gaussian Modeling from OpenVDB Leaf nodes}
Our Gaussian model is generated from data stored in the OpenVDB grid nodes and hence a primary understanding of the layout is essential before re-modeling of the data. The grid spatial structure is infact a deciding factor for clustering of data in different LOD settings of our scheme. 

\begin{figure}[H]
	\centering 
	\includegraphics[width=0.9\columnwidth, alt={Optix Shader pipeline used for rendering 3D Gaussians.}]{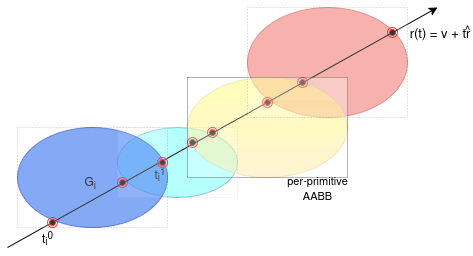}
	\caption{%
		2D-Schematic diagram showing how a ray intersects with Gaussians, that are each placed inside of an AABB for hardware accelerated BVH traversal in OptiX.
	}
	\label{fig:ray-gauss-intersection}
\end{figure}

\subsection{Grid Layout}

OpenVDB provides a data format(.vdb) and a library of accessor functions, iterators and data utility functions for that format. It is primarily used for storing and accessing volumetric data and was developed mainly to reduce storage for sparse volumes through a hierarchical, axis-aligned grid tree composed of regularly spaced voxels with configurable active and inactive sub-trees. The grid is characterized by a tree structure of fixed depth. The origin or root of the tree is called the \textbf{Root Node}, which spans the entire volume and subdivides into multiple internal nodes of the top-most level. These top-level nodes are disjoint as per our knowledge, $2^n$ cubical regions in 3D space, where n corresponds to the height of the tree. Each node is identified by a unique origin, providing a deterministic access coordinates of the data within each node in \textit{index space}(For details on index spaces in OpenVDB and their influence on instance transformations please refer to our earlier work\cite{Sharma2025GaussianParticleApproximation}). The number of such nodes depends to the extent of the dataset, however their size remains static. Thus, simplifying memory allocation and facilitating dense packing of active regions.

We focus on the commonly used \textbf{5-4-3} grid configuration, also used in our Gaussian modeling settings. In this layout:
\begin{itemize}
	\item Each top-level(``5-level'') node has a resolution of $32^3 = 32768$ child nodes.
	\item Each child(``4-level'' node) contains $16^3 = 4096$ further child nodes.
	\item The lowest level(``3-level'' or leaf nodes) contains $8^3 = 512$ voxels, which store the actual scalar values.
\end{itemize}

To support sparsity, OpenVDB avoids storing inactive voxels. For example, a fully populated $4096^3$ grid(within a single top-level node) can consume over 128 GB using only half-precision floats. Instead, each node uses the following type of bitmasks to track activity and storage:
\begin{itemize}
	\item \textbf{Child Mask:} A 1-bit flag per child indicates whether its subtree contains any active voxels.
	\item \textbf{Value Mask(Tile):} Also 1 bit per child, this indicates whether a subtree contains a constant value, avoiding allocation of deeper nodes. These regions are known as \emph{tiles}.
\end{itemize}

\subsection{3D Volumetric Gaussians generation}
In this work, we use leaf Nodes as our spatial anchors for processing of voxels and where each Gaussian gets initialized to a distinct and non-intersecting voxel subregion of the leaf Node. This results in a reduced-fidelity approximation of the original volume, as each voxels-subregion is encoded using a spherical or ellipsoidal Gaussian. Using the OpenVDB grid tree iterators we can traverse all leaf nodes and further traverse data within each leaf using leaf iterators. Converting a voxels-subregion into Gaussians at varying resolutions, depending on the level of detail(LOD), involve a number of steps as described in the subsection below. For each Gaussian in our model, we extract the following key attributes:

\begin{enumerate}
	\item \textbf{Position}: The center of the Gaussian in world space, typically computed as the centroid of a voxel region or bounding box.
	\item \textbf{Opacity}: A value representing the average scalar value of voxels within the region, used to preserve local material properties.
	\item \textbf{Covariance Vector}: The diagonal elements of a $3\times3$ covariance matrix representing the Gaussian's spatial extent along each axis. Since we currently use axis-aligned Gaussians, only the diagonal is stored. For isotropic Gaussians, this reduces to a single scalar value(although a voxel can have different resolution in each dimension). This representation supports both spherical and ellipsoidal Gaussian shapes.
\end{enumerate}

Before processing the leaf Nodes, it is necessary to check whether the grid uses \textbf{tiled} storage. Tiling optimizes memory usage by representing uniform regions with a single value and a value mask at an intermediate node level, omitting the corresponding leaf Node altogether. While tiles can simplify the detection of homogeneous regions and enable conversion into single Gaussians, they require explicit traversal of the grid tree, which is more computationally expensive than directly accessing leaf nodes at a fixed depth, nevertheless it is important to check for them for accurate fitting. Tiles may exist at both intermediate and root levels in the grid hierarchy.

\textbf{Leaf Nodes} can be either dense or sparse, depending on whether all voxels within the Node contain valid data. Regardless of density, each leaf node has a fixed resolution of $512$ voxels, which makes them well-suited for parallel processing. For parallelism on the CPU, OpenVDB relies on Intel OneAPI Thread Building Blocks(TBB)~\cite{intel_onetbb}, which is a required dependency for performing operations such as leaf node iteration, grid construction, transformation, filtering, resampling etc. TBB is not shipped by OpenVDB itself, but it is the default backend for task-based parallelism for these operations. Our parallelization strategy assigns one leaf Node per thread. Each thread collects Gaussian data in local buffers and writes the results to a global output array using a mutex for thread safety. Each thread maintains the following data structures:

\begin{itemize}
	\item Thread-local buffers for storing the properties of each fitted Gaussian, including its 3D position, scalar opacity, and diagonal elements of the covariance matrix. Rotational information is not stored, since our Gaussians are axis-aligned.
	\item A metadata struct for storing axis-aligned bounding boxes per Gaussian and indexing information such as offsets and counts in the global output.
\end{itemize}

\subsection{Fitting Gaussians to leaf Nodes}
We use different strategies for converting dense and sparse OpenVDB leaf Nodes into single or a set of Gaussians.
\subsubsection{Dense leaf Nodes}
A constant size block-based approach is used for dense Nodes. The Node is partitioned into regular $2\times2\times2$, $4\times4\times4$ voxels, or complete leaf Node is used as one block($8\times8\times8$ voxels). This is based on a user-input value for the level-of-detail(LOD) selection, and each block is fitted with a single Gaussian. The Gaussians can be isotropic if voxel resolution is same in each dimension. The Gaussian position is computed as the center of the voxel block-group and is transformed into world space by applying the grid transform to the index space center coordinates. The covariance diagonal is estimated using half the block size scaled by the voxel resolution, resulting in axis-aligned radii in all three dimensions. 

\subsubsection{Sparse Leaf Nodes}
\begin{itemize}
	\item \textit{Smart grouping.} For non-dense Nodes, voxel occupancy is irregular and hence a more adaptive strategy is implemented. Our algorithm scans all active voxels in ZYX order(according to OpenVDB’s memory layout), and attempts to group them into blocks of $2\times2\times2$ voxels. If all eight voxels of a block are present and unused, they are grouped and one Gaussian is fitted in this region.
	For a partially filled block or a sparse leaf Node from which no blocks can be formed, a fallback grouping is done. Adjacent voxel pairs are identified based face adjacency along X, Y, or Z and formed into a group. Any remaining unpaired voxels are added as singletons. This two-level grouping ensures that as much local structure as possible is captured in larger aggregates, while still preserving all voxel information.
	
	\item \textit{Strict blocks + single voxels.} This strategy follows greedy grouping of $2\times2\times2$ voxels, if available. And all ungrouped active voxels are directly converted into single-voxel Gaussians, each centered at the voxel's center position and assigned a covariance diagonal vector based on the voxel sizes in each dimension. This can generate a lot of Gaussians based on structure and quantity of sparse leaf Nodes in a dataset but its simplicity makes it faster in fitting. 
	
	\item \textit{Single biggest Gaussian per leaf.} In this strategy, all active voxels within a sparse leaf node are approximated using a single Gaussian. The position is computed as the centroid of the \textit{minimum} bounding box, which is computed incrementally, by expanding a singular axis-aligned bounding box(AABB) while traversing every active voxel coordinates in each dimension using the leaf Node iterator. The covariance diagonal vector is then derived from the dimensions of this bounding box. This method is particularly effective when preserving fine detail is less critical and a compact representation is the main priority.
\end{itemize}

\subsubsection{Gaussian Construction and per-Gaussian AABB}
For each identified group: whether from dense or sparse regions, the group’s centroid defines the Gaussian position which are transformed into world space using the grid’s transform, and the covariance is computed from the bounding box size scaled by the voxel size. Radii are always clamped to a minimum, to avoid degenerate ellipsoids, and opacity is taken as the average scalar value of the group. Each Gaussian is filtered by its average opacity: only groups whose average or individual opacity exceeds a small user-defined threshold($10^{-3}$ - $10^{-6}$) are retained. This prevents unnecessary allocation of Gaussians in near-empty regions. Finally, the resulting Gaussian parameters, i.e. position, covariance, opacity and metadata are written to output buffers. This flexible and multi-modal approach allows for adaptive fitting of Gaussians depending on the density and structure of the voxel data, balancing between compactness and accuracy.

Later for rendering Gaussians in OptiX, each primitive must be enclosed within an axis-aligned bounding box(AABB), which forms the basis of the bounding volume hierarchy(BVH) used for hardware-accelerated ray tracing. The AABB is centered at the Gaussian’s center position and extends symmetrically along each axis based on the Gaussian’s covariance. To control the size of the bounding box relative to the Gaussian's spread, a ``sigma'' factor can be applied to each of the three radii. $\sigma = 1$, $2$, or $3$ determine how much of the Gaussian distribution is enclosed, trading off between tight bounds and conservative coverage. The resulting bounds are also clamped to the entire scene's bounding box to ensure they remain valid. These AABBs are used to populate the OptiX Geometry Acceleration Structure(GAS), enabling high-performance ray-Gaussian intersection and efficient culling during rendering.

\subsubsection{Level of Detail(LOD) Control}
As previously described, both dense and sparse leaf Nodes can be converted into Gaussian representations using different strategies. Every combination of a dense block-group and a non-dense strategy effectively is a LOD. Since our LOD configurations are controlled by the granularity of voxel grouping, it can be adjusted depending on application requirements. For dense regions, fixed-size blocks(e.g., $2^3$ or $4^3$) allow for fine to coarse approximations, with larger blocks yielding fewer, more spatially extensive Gaussians. In sparse regions, we have adaptive methods such as ones having a fallback mechanism for forming smaller groups out of distributed Gaussians or adding rogue singletons as-is for preserving structure, and finally the coarsest approximations of collapsing an entire leaf node into a single Gaussian to provide fast approximations when details are not critical. By selecting an appropriate combination of these strategies, users can balance fidelity and compactness to achieve scalable and resolution-adaptive representations of volumetric data. For details on additional ways of clustering voxels for Gaussian generation please refer to Section 4 of our earlier work\cite{Sharma2025GaussianParticleApproximation}. It provides details on variance-based Intra-leaf Voxels clustering, Inter-leaf node clustering and Grid Tree mirroring.

\section{Rendering Pipeline: Density Visualization}
Our renderer visualizes the 3D anisotropic Gaussians generated from an OpenVDB volumetric dataset, using a purely absorption-based model currently because we are evaluating it for Scientific Visualization purposes. There are no surfaces, no emission, and no scattering. The color or "radiance" reaching the camera is derived entirely from a transfer function color mapping of the accumulated optical depth along each viewing ray. The goal here is to expose the structure of volumetric data rather than simulate realistic lighting.

\subsection{Optical Depth and Final Color}
For each camera ray \( r(t) = v + t\hat{r} \), where $v$ is the ray origin and $t$ is the distance along the ray from the origin and $\hat{r}$ is the unit vector in the direction of the ray, we compute the total optical depth \( T \) by accumulating contributions from all intersected 3D anisotropic Gaussians:

\begin{equation}
	T = \sum_{i \in \mathcal{I}} \tau_i = \sum_{i \in \mathcal{I}} c_i \int_{t_i^0}^{t_i^1} \rho_i(r(t)) \, dt
	\label{eq:optical_depth}
\end{equation}

where:
\begin{itemize}[noitemsep]
	\item \( \rho_i(x) \) is the unnormalized Gaussian density centered at \( \mu_i \) with covariance matrix \( \Sigma_i \), where the value of $x$ in this context is $r(t)$
	\item \( [t_i^0, t_i^1] \) defines the segment of the ray overlapping Gaussian \( i \),
	\item \( c_i \) is the scalar opacity coefficient,
	\item \( \tau_i \) is the optical depth contribution from Gaussian \( i \),
	\item \( \mathcal{I} \) is the set of Gaussians intersected by the ray.
\end{itemize}

To determine whether a ray intersects a Gaussian, we solve a quadratic equation derived by substituting a fixed density threshold \( \kappa \) in the Gaussian density falloff equation. Then, each Gaussian's spatial support is defined as the region where its unnormalized density satisfies:
\[
\rho_i(x) = \exp\left(-(x - \mu_i)^\top \Sigma_i^{-1}(x - \mu_i)\right) \geq \kappa
\]

Taking the logarithm of both sides yields the squared Mahalanobis distance condition, which is a quadratic form for establishing a non-outlier region:
\[
(x - \mu_i)^\top \Sigma_i^{-1}(x - \mu_i) \leq -\log \kappa
\]

Substituting the ray \( r(t) = v + t\hat{r} \) into this inequality, then rearranging and expanding this leads to a quadratic in \( t \):
\begin{equation}
C t^2 + B t + A + \log \kappa \leq 0
\label{eq:quadratic_in_t}
\end{equation}
where the coefficients A, B and C can be computed as described in the Appendix A. 
This defines a boundary for the spatial support of a Gaussian and solving this quadratic yields at most two real roots \( t_i^0 \) and \( t_i^1 \), corresponding to the entry and exit distances along the ray where it intersects this ellipsoidal support of the Gaussian. These bounds are then used to compute the Gaussian's contribution to the optical depth by evaluating the line integral\cite{inproceedings} of the density function along the ray segment(Appendix B):

\begin{equation}
	\tau_i = c_i \int_{t_i^0}^{t_i^1} \rho_i(r(t)) \, dt
	\label{eq:ray_gaussian_integral}
\end{equation}

The total optical depth \( T \) is then converted into a visibility value using the exponential attenuation model:

\begin{equation}
	V = 1 - \exp(-T)
	\label{eq:visibility}
\end{equation}

Finally, the visibility is mapped to a color value via a user-defined transfer function:

\begin{equation}
	L =
	\begin{cases}
		\text{TransferFunction}(V), & \text{if } T > 0 \\
		L_{\text{bg}}, & \text{otherwise}
	\end{cases}
	\label{eq:final_output}
\end{equation}

As in standard ray marching renderer for Volumes, the transfer-function is applied only after optical depth accumulation and is otherwise decoupled from the data model. Consequently, our method allows dynamic transfer function updates at runtime without requiring any reinitialization of the Gaussian representation.

\begin{figure}[H]
	\centering 
	\includegraphics[width=0.9\columnwidth, alt={Optix Shader pipeline used for rendering 3D Gaussians.}]{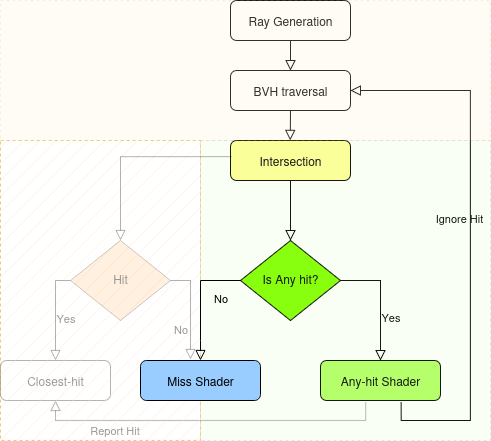}
	\caption{%
		OptiX shader workflow used in our Gaussians renderer.%
	}
	\label{fig:optix_workflow}
\end{figure}

\subsection{GPU-Based Ray Marching of Gaussians}

In a nutshell, for each pixel the GPU ray generation kernel performs the following steps:

\begin{enumerate}[noitemsep]
	\item \textbf{Ray Setup:} Generate a camera ray and find its interval of intersection with the bounding box of our Gaussian model of the volume.
	\item \textbf{Intersection Collection:} Traverse the acceleration structure to populate list of all Gaussians intersected by the ray.
	\item \textbf{Sorting:} Sort Gaussians by entry distance to enable front-to-back accumulation.
	\item \textbf{Optical Depth Integration:} Compute and accumulate optical depth contributions from each intersected Gaussian.
	\item \textbf{Color Mapping:} Compute visibility and apply a transfer function to obtain the final pixel color.
	\item \textbf{Fallback:} If no Gaussians are hit, output the background color \( L_{\text{bg}} \).
\end{enumerate}

Our 3D Gaussian renderer is implemented in CUDA and uses NVIDIA's OptiX8.1 ray tracing framework for fast BVH traversal and ray-primitive intersection queries. The programmable pipeline for defining custom ray traversal and shading behavior in our renderer primarily uses the following functions: ray generation, intersection, and anyhit shaders to collect contributions along the ray from all intersected Gaussians, while the miss shader assigns the background color for rays that do not intersect any primitives. The overall structure of the OptiX pipeline as used in our method is illustrated in Figure~\ref{fig:optix_workflow}.

 In the \textit{\_\_raygen\_\_} program(short for ray generation), a primary ray \( r(t) = v + t\hat{r} \) is launched for each pixel from the camera position \( v \), which is first clipped against the AABB of the Gaussian model of the volume, to restrict processing to the region containing valid data. The ray is then traced using a per-ray payload structure(\texttt{prd}) that stores all intersected Gaussians along the path.

Each volumetric Gaussian \( \rho_i(x) \), defined by its center \( \mu_i \), opacity \( c_i \), and covariance matrix \( \Sigma_i \), is represented as a custom primitive enclosed within an axis-aligned bounding box(AABB) as illustrated in Figure~\ref{fig:ray-gauss-intersection}. These AABBs are required for building acceleration structures and enable efficient BVH-based traversal. The size of each AABB is determined by the Gaussian's ellipsoidal support, which is defined via a fixed density threshold \( \kappa \). Specifically, the support region corresponds to the Mahalanobis level set at which the unnormalized Gaussian density falls to \( \kappa \), enclosing most of the meaningful contribution while culling the long tails of the function.

In the \textit{\_\_intersection\_\_} shader, the ray-Gaussian intersection is handled analytically using the Equation~\ref{eq:quadratic_in_t}, where the ray is tested for overlap with the ellipsoidal support. If the ray intersects the support region, the entry and exit distances along the ray are computed. These values, along with the Gaussian primitive index, are then passed to the any-hit program from the intersection shader using attribute variables.

In the \textit{\_\_anyhit\_\_} shader, each valid hit is appended to a per-ray buffer of Gaussian particles(\texttt{prd.particles[tail]}, where tail corresponds to count of particles) and the relevant counter is incremented. We use \texttt{optixIgnoreIntersection()} to allow continued traversal, enabling accumulation of overlapping contributions along the ray.

Once the traversal concludes and all intersections have been reported, execution goes back to the raygen program which receives the full list of intersected Gaussians. These are then sorted by their entry distance \( t_i^0 \) to ensure correct front-to-back compositing. For each intersected Gaussian, the ray segment bounded by \( [t_i^0, t_i^1] \) is used to compute the integral contribution to optical depth, as defined in Equation~\eqref{eq:ray_gaussian_integral}. The total optical depth \( T \) is computed as in Equation~\eqref{eq:optical_depth}, and visibility is then evaluated using the exponential attenuation model in Equation~\eqref{eq:visibility}. Finally, the visibility is mapped to a color value via the transfer function, as shown in Equation~\eqref{eq:final_output}.

This pipeline computes Gaussian contributions analytically by tracing rays through the scene and evaluating the line integral of each ray segment intersecting a Gaussian. Unlike rasterization-based splatting, which emphasizes view-dependent surface projections, our approach models the full 3D volumetric extent, enabling higher-fidelity reconstruction of the underlying data.

\subsection{Rendering Algorithm}

The key steps of the rendering process, combining the optical depth formulation with the GPU pipeline can hence be highlighted as follows:

\medskip

\noindent
\textbf{Initialize:} \( T \leftarrow 0 \)

\noindent
\textbf{For each pixel:}
\begin{itemize}
	\item Generate ray \( r(t) = v + t\hat{r} \) from camera
	\item Compute bounds of intersection of the  ray with the Gaussian model of the Volume \( [t_{\text{enter}}, t_{\text{exit}}] \)
	\item Invoke \texttt{traceRay()} with payload \texttt{prd}
\end{itemize}

\noindent
\textbf{In \texttt{\_\_intersection\_\_gaussian}:}
\begin{itemize}
	\item Solve the quadratic equation derived from the Mahalanobis condition applied to the ray:
	\[
	(r(t) - \mu_i)^\top \Sigma_i^{-1}(r(t) - \mu_i) = -\log(\kappa)
	\]
	This yields entry and exit points \( t_i^0 \), \( t_i^1 \) for the ray-Gaussian intersection.

	\item If intersected, compute entry and exit points \( t_i^0 \), \( t_i^1 \)
	\item Pass \((t_i^0, t_i^1, i) \) to \texttt{\_\_anyhit\_\_}
\end{itemize}

\noindent
\textbf{In \texttt{\_\_anyhit\_\_radiance\_\_gaussian}:}
\begin{itemize}
	\item Append hit to \texttt{prd.particles[tail]} and increment tail
	\item Call \texttt{optixIgnoreIntersection()} to continue traversal
\end{itemize}

\noindent
\textbf{In \texttt{raygen} after traversal:}
\begin{itemize}
	\item Sort all intersected Gaussians by entry distance \( t_i^0 \)
	\item For each Gaussian \( i \in \mathcal{I} \):
	\begin{itemize}
		\item Retrieve \( \mu_i \), \( \Sigma_i \), \( c_i \)
		\item Compute optical depth contribution \( \tau_i \) via ray-Gaussian line integral
		\item Accumulate: \( T \leftarrow T + \tau_i \)
		\item \textbf{If} \( T > 0.999 \) or \( T < 0.001 \), \textbf{break}
	\end{itemize}
	\item Compute visibility: \( V = 1 - \exp(-T) \)
	\item Map \( V \) to final color using the colormap
	\item \textbf{If} no Gaussians were hit: assign background color \( L = L_{\text{bg}} \)
\end{itemize}

\begin{figure}[h]
	\centering
	
	\begin{subfigure}{0.30\columnwidth}
		\includegraphics[width=\linewidth]{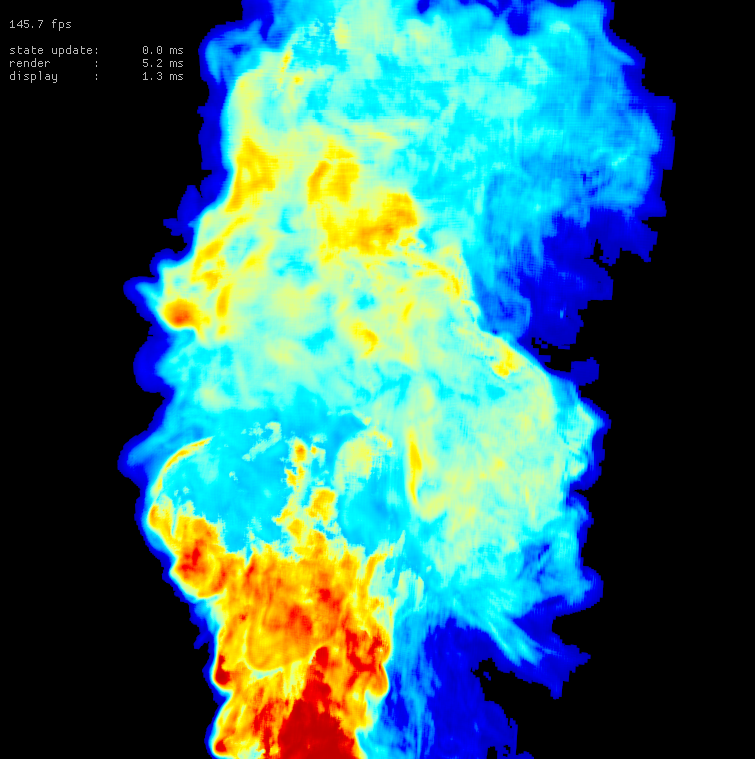}
	\end{subfigure}
	\hfill
	\begin{subfigure}{0.30\columnwidth}
		\includegraphics[width=\linewidth]{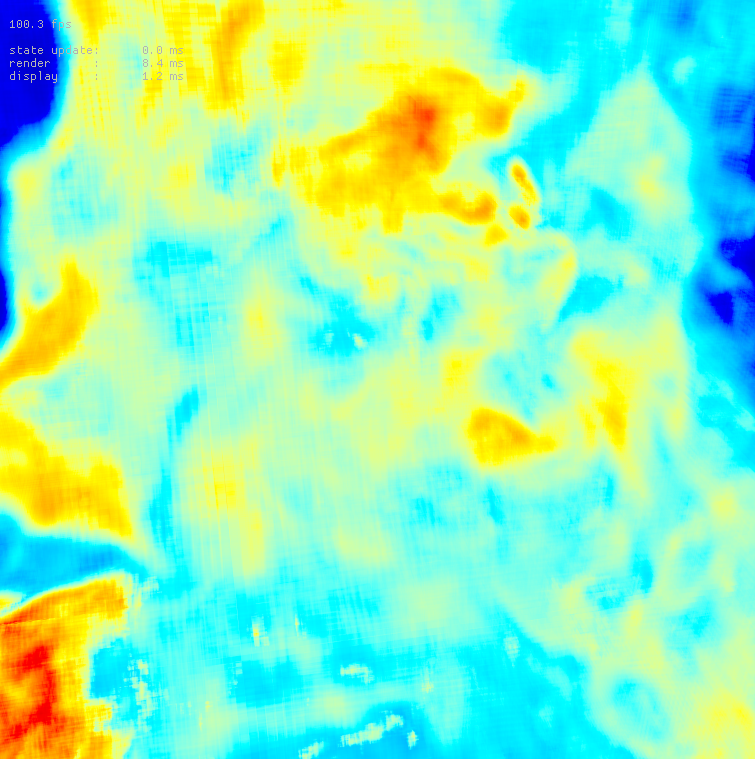}
	\end{subfigure}
	\hfill
	\begin{subfigure}{0.30\columnwidth}
		\includegraphics[width=\linewidth]{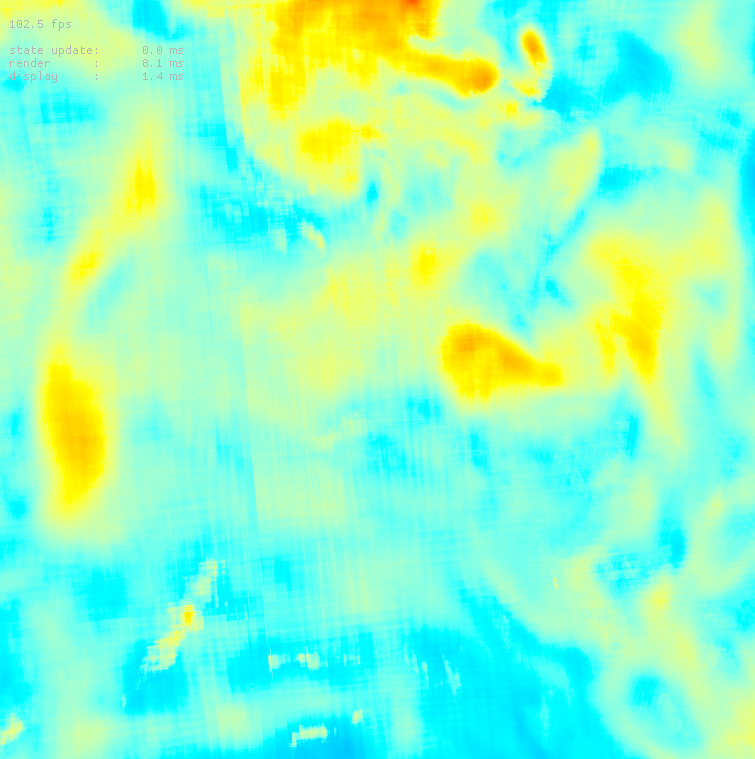}
	\end{subfigure}
	
	\vspace{1em} 
	
	\begin{subfigure}{0.30\columnwidth}
		\includegraphics[width=\linewidth]{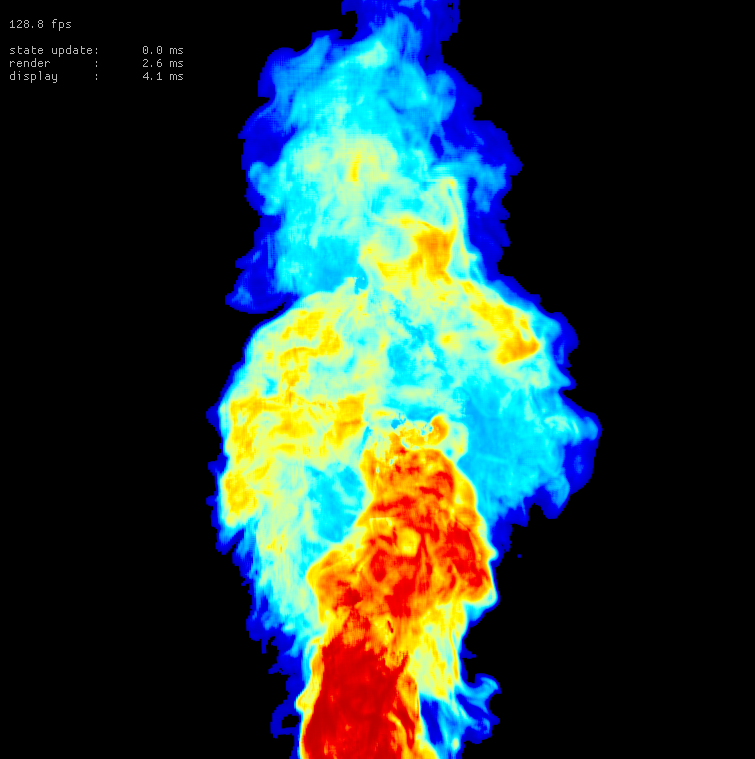}
	\end{subfigure}
	\hfill
	\begin{subfigure}{0.30\columnwidth}
		\includegraphics[width=\linewidth]{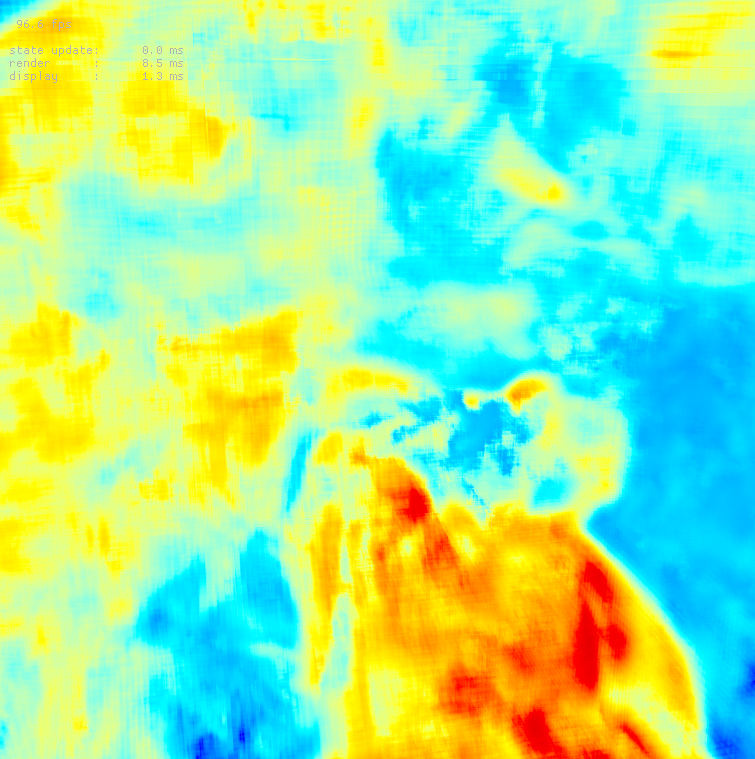}
	\end{subfigure}
	\hfill
	\begin{subfigure}{0.30\columnwidth}
		\includegraphics[width=\linewidth]{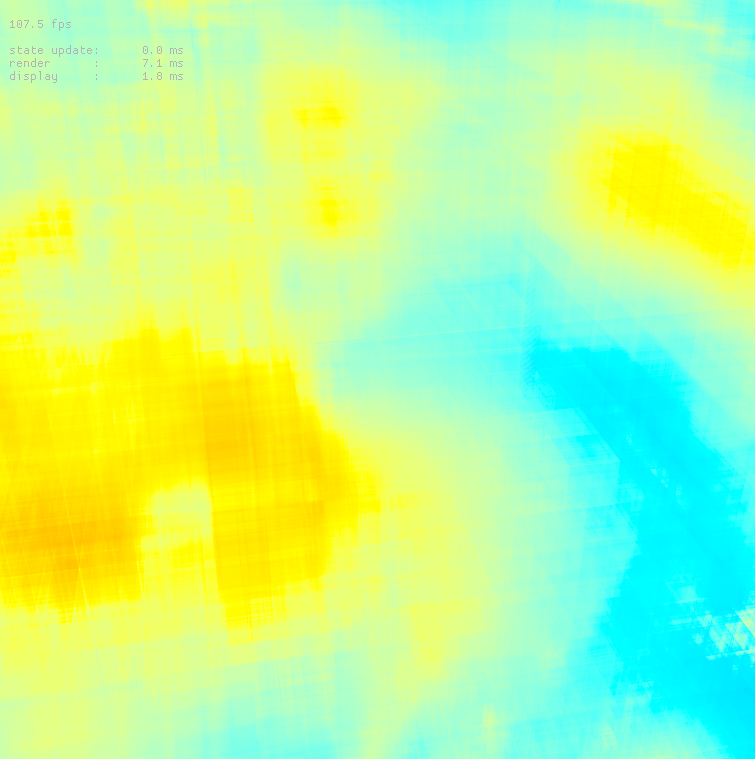}
	\end{subfigure}
	
	\caption{%
		Smoke2 dataset zoomed-in in our interactive renderer from two different camera view angles showcasing how the Gaussians model the internal structure of the volume. 
	}
	\label{fig:zoom_smoke2}
\end{figure}

\subsection{Application and System}
We implement two custom renderers with interactive viewers in C++ and CUDA with OptiX8.1, leveraging GLFW for real-time user interaction through keyboard and mouse input, with a trackball-style camera for zoom, pan, and rotation. The Gaussian Renderer has been described already in previous subsections.
 
\begin{figure}[tb]
	\centering
	
	\begin{subfigure}{0.48\columnwidth}
		\includegraphics[width=\linewidth]{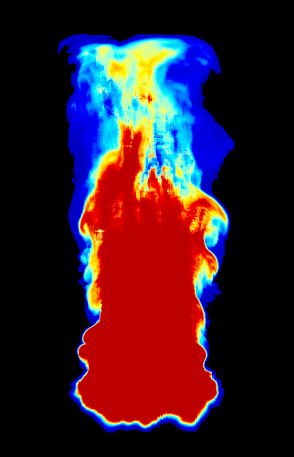}
	\end{subfigure}
	\hfill
	\begin{subfigure}{0.48\columnwidth}
		\includegraphics[width=\linewidth]{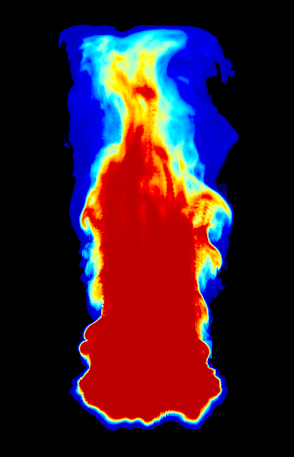}
	\end{subfigure}
	
	\begin{subfigure}{0.48\columnwidth}
		\includegraphics[width=4.24cm, height=4.13cm]{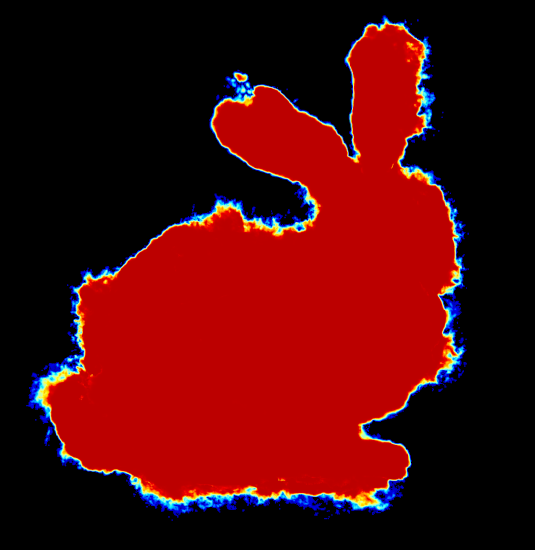}
	\end{subfigure}
	\hfill
	\begin{subfigure}{0.48\columnwidth}
		\includegraphics[width=\linewidth]{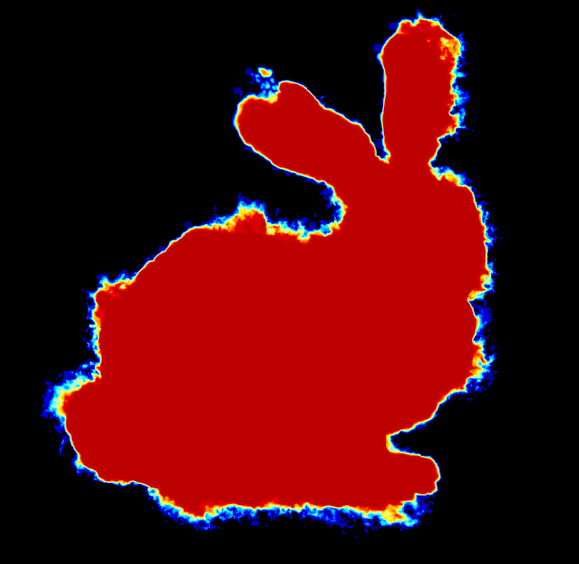}
	\end{subfigure}
	
	\caption{Comparison of Gaussian-rendered(left) and SciVis(right) versions for fire and bunny\_cloud datasets.}
	\label{fig:openvdb_fire_bunny}
\end{figure}

The second renderer is the \textbf{SciVis renderer}, which performs volume ray marching using primary-ray casting and similarly models only absorption, however using a different ray marching and optical depth calculation scheme. It initializes camera rays in \textit{\_\_raygen\_\_} in the same way as our Gaussians renderer, but instead of following our custom ray marching logic, it uses NanoVDB’s HDDA(Hierarchical Digital Differential Analyzer)\cite{Museth2014HDDA} traversal to efficiently accumulate optical depth along the ray by integrating sampled density values between the hit points through the voxels. Its core ray marching logic is implemented in the \textit{\_\_closesthit\_\_} shader instead of the \textit{\_\_anyhit\_\_} shader that is used in the Gaussians renderer because we need to report only the closest hit to the HDDA function which then accumulates the optical depth till the exit point of the volume. Optical depth is then converted in a similar fashion as the Gaussians renderer using transfer functions. In Figure~\ref{fig:openvdb_fire_bunny}, two of the sample OpenVDB datasets for volumes are shown as rendered by our methods for Gaussians rendering and SciVis rendering. Figure~\ref{fig:zoom_smoke2} showcases how the interactive controls of the renderer allow us to explore the volumes internally by zooming in.

The SciVis renderer implements the closest external reference implementation of transmittance accumulation for OpenVDB datasets with primary ray-casting for absorption modeling, hence we use it as our ground truth. Our hardware setup includes an NVIDIA GeForce RTX 4090 Laptop GPU and an Intel Core i9-14900HX CPU. This combination is crucial, as OpenVDB relies on multi-threaded CPU processing for voxel and node management, while OptiX operates entirely on the GPU. Prior to rendering, all Gaussian parameter buffers are transferred from host to device memory as separate GPU buffers to enable efficient hardware-accelerated processing. For a detailed explanation on managing and copying data buffers to device memory, as well as setting up the Shader Binding Table(SBT), Geometry Acceleration Structure(GAS), and Instance Acceleration Structure(IAS), i.e. all essential components for utilizing OptiX’s BVH traversal and ray tracing pipeline please refer to Section~5 of our previous work on Gaussian rendering~\cite{Sharma2025GaussianParticleApproximation}.

\section{Unstructured Point Clouds}
It is apparent that the Gaussian modeling and rendering methods that we have proposed can support unstructured point clouds in addition to grid-based volumes. For this, we first convert a point cloud or particle dataset to an object of OpenVDB \textit{PointDataGrid} class type which can efficiently encode these points in a sparse hierarchical grid. For an illustrative example, we are using the 1million particle ``Dark Sky'' cosmological dataset\cite{NVIDIA2022DarkSky1M}, originally provided as a plain-text \texttt{.xyz} file containing 3D positions and velocity vector for each particle. The .xyz file is parsed and an OpenVDB PointDataGrid is constructed from this list of particles using the function \texttt{openvdb::points::createPointDataGrid}, where each point contains a 3D position and a scalar velocity magnitude derived from the velocity vector. The PointDataGrid is a sparse voxelized point cloud where the particle positions are stored in the built-in \texttt{"P"} attribute the velocity magnitude is stored in a  custom per-point attribute named \texttt{"velocity"}. 

To achieve this, the 3D positions and velocities are first extracted into separate buffers. A voxel size is then computed to ensure approximately one particle per voxel, which aids in efficient spatial indexing. The PointDataGrid object is then created using first the positions buffer, the voxel size and a corresponding linear index-to-world transform. After the grid is initialized, the \texttt{"velocity"} attribute has to be manually added to all points. It can be filled by iterating over the leaf nodes and their constituent points using leaf and index iterators. Using the write handle to each node’s \texttt{"velocity"} attribute array, we can populate it with the values from the velocity vector. 

\begin{figure}[h]
	\centering
	\captionsetup[subfigure]{skip=0pt}
	\begin{subfigure}[b]{\columnwidth}
		\centering
		\includegraphics[width=\textwidth]{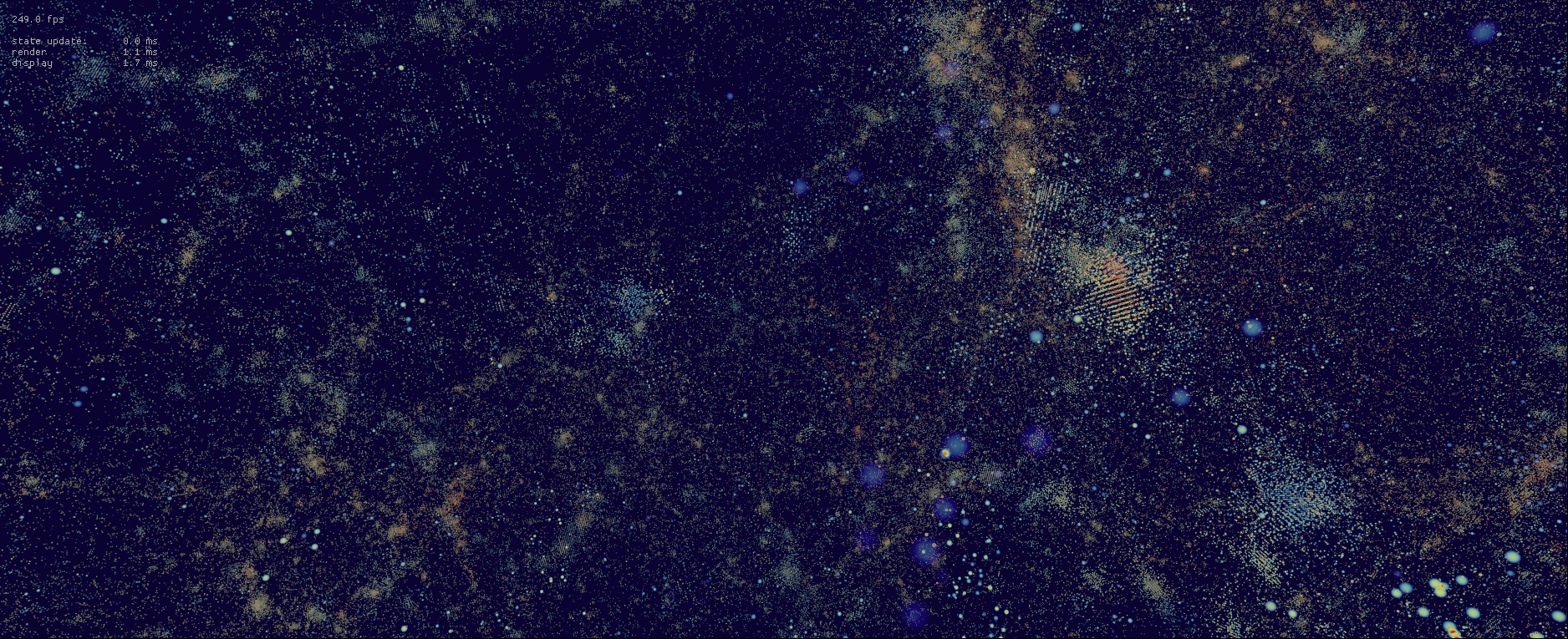}
	\end{subfigure}
	
	\begin{subfigure}[b]{\columnwidth}
		\centering
		\includegraphics[width=\textwidth]{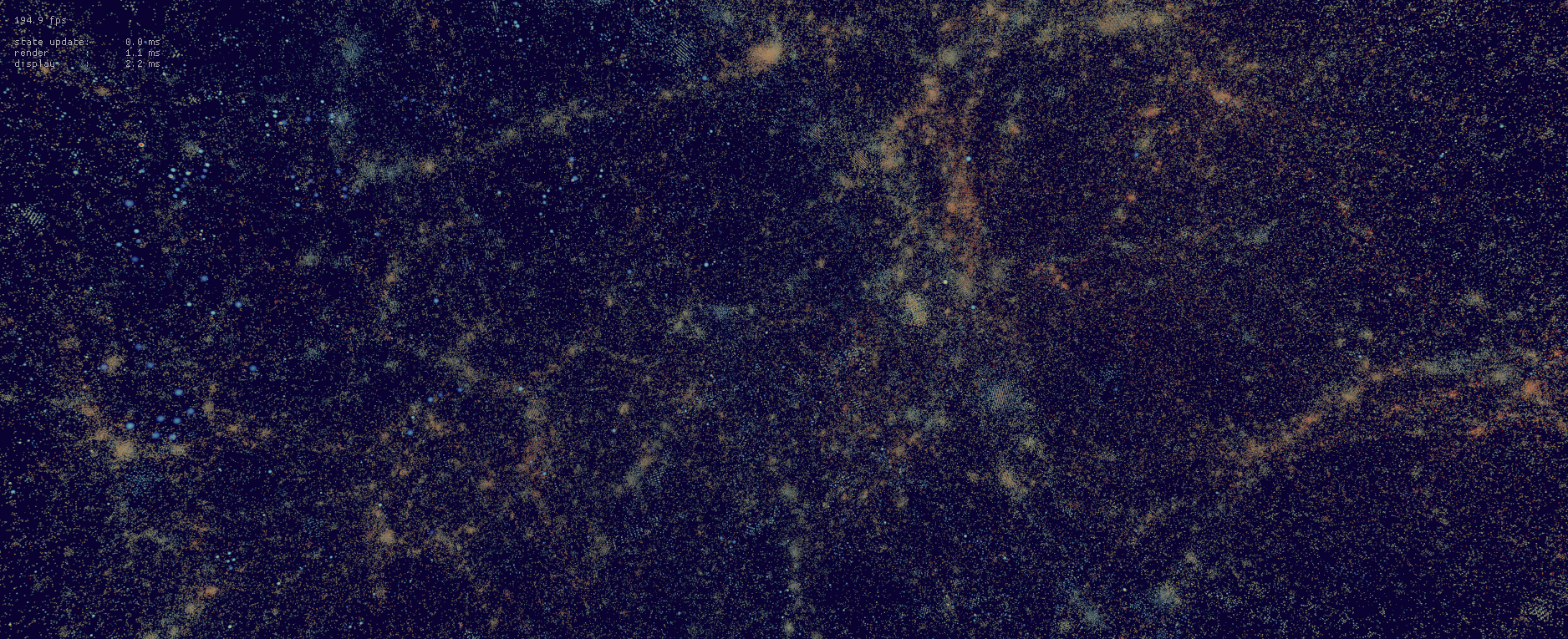}
	\end{subfigure}
	
	\begin{subfigure}[b]{\columnwidth}
		\centering
		\includegraphics[width=\textwidth]{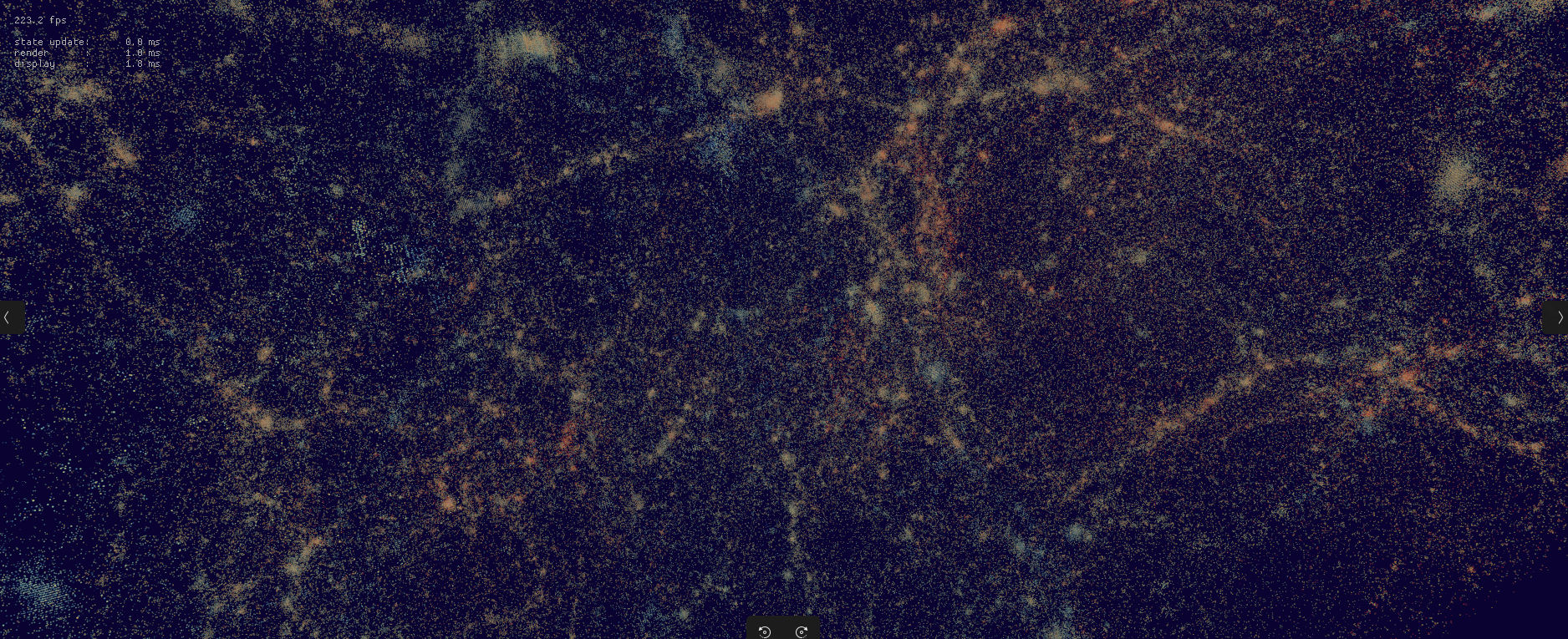}
	\end{subfigure}
	
	\begin{subfigure}[b]{\columnwidth}
		\centering
		\includegraphics[width=\textwidth]{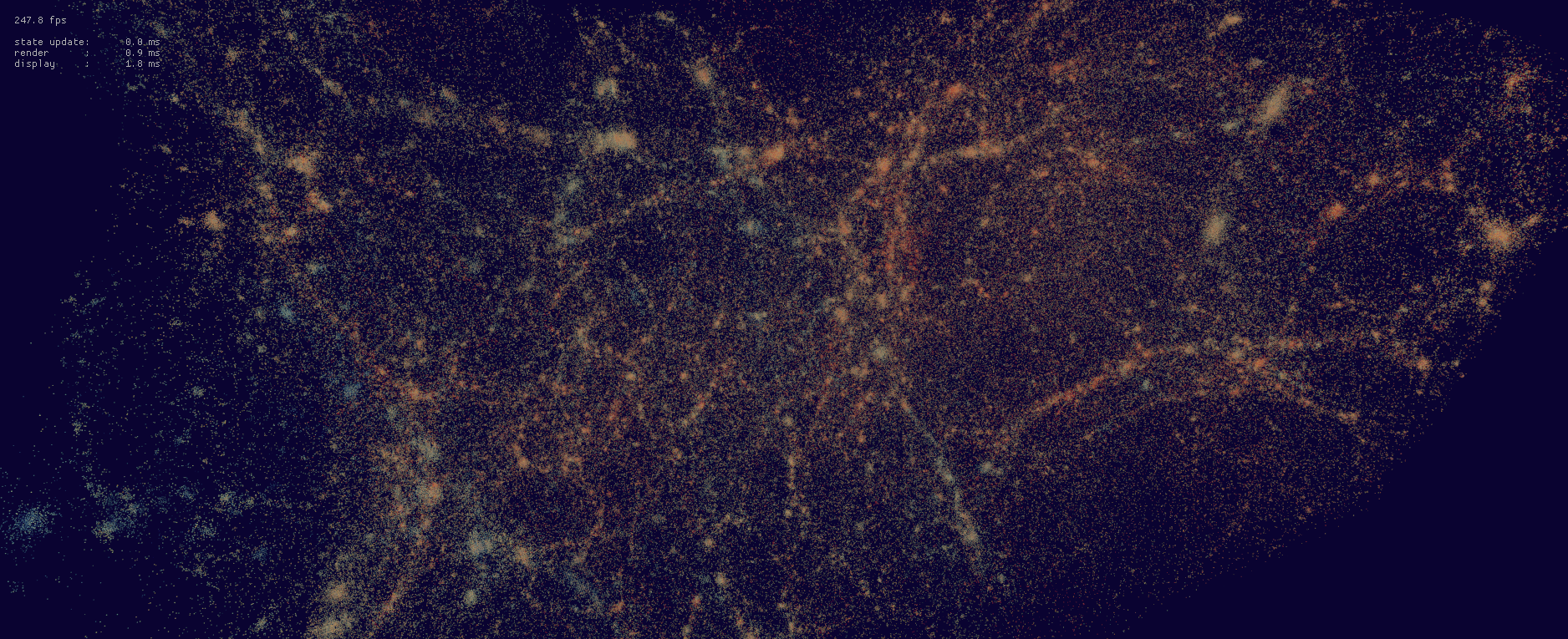}
	\end{subfigure}
	
	\caption{The \texttt{dark\_sky} dataset rendered using the Jet transfer function. From top to bottom, a progressive zoom-out reveals the effect of increasing spatial density: as more Gaussian particles accumulate within the view, the rendering shifts from cooler blue hues to warmer reds and oranges, indicating higher opacity. In the topmost image, individual Gaussians are still discernible and appear as soft, low-opacity blue regions.
	}
	\label{fig:dark_sky_zoom_out}
\end{figure}

The resulting PointDataGrid can be efficiently queried using the OpenVDB point iterators during Gaussian modeling. Once converted to 3D Gaussians the rest of the rendering pipeline works without requiring any modifications. While the visual appearance can be customized through different transfer functions and shading parameters such as blending the background color or not, each particle is rendered as a soft, isotropic Gaussian, with its velocity magnitude encoded as the opacity attribute. The resulting rendering can be seen in Figure~\ref{fig:dark_sky_zoom_out} as progressive zoom out to reveal high density regions in the dataset. 

\section{AMR Volumes}
Adaptive Mesh Refinement(AMR) is a widely used technique across scientific computing domains, including astrophysics, fluid dynamics, and climate modeling, to facilitate finer resolution in regions of interest while minimizing memory and computation in feature-deficient regions. However, AMR data structures are typically composed of nested, non-uniform rectilinear grids that are challenging to render directly, particularly in GPU-based ray tracing pipelines that assume uniform grid domains.

For supporting interactive visualization of AMR volumes in our renderer, each refinement level should be converted into its own OpenVDB FloatGrid object. Refinement masks can be used for indicating regions that are further refined. The mask enables hierarchical compositing by masking lower-resolution data in regions occupied by higher-resolution levels. Each level can be assigned a voxel size proportional to its refinement level and transformed into it's own world-space coordinates instead of frame-of-reference of the base level. This ensures that nested levels align precisely in space, even when rendering in a vertex-centered configuration with voxel data. To avoid interpolation artifacts at refinement boundaries, the mask clips lower-resolution contributions. For more details on this process please read this master thesis that rendered AMR to VDB converted grids in Houdini\cite{Ravindran2017Houdini}.

\begin{figure}[tp]
	\centering 
	\includegraphics[width=0.5\columnwidth, alt={Optix Shader pipeline used for rendering 3D Gaussians.}]{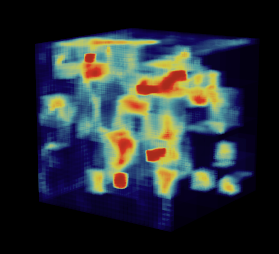}
	\caption{Enzo Tiny Cosmology 3-grids VDB dataset rendered with Gaussians}
	\label{fig:tiny_enzo}
\end{figure}

For an illustrative example, we have used the VDB files corresponding to three levels of refinement of the Enzo Tiny Cosmology sample dataset provided at ytini\citen{ytini_amr}. Input to our Gaussian generation and rendering pipeline is hence an array of files, each loaded as a separate OpenVDB grid complete with its own voxel size and world transform for placing it correctly in the world space. These are then seamlessly converted into Gaussians each with its correct world position determined using the respective transformation of the Grid. This combination of sparse hierarchical storage and Gaussian representation enables scalable and visually consistent rendering of multiscale datasets. Figure~\ref{fig:tiny_enzo} shows rendering of this dataset with jet transfer function interpolated with background color. Finer resolution data in shown in red and warmer hues. 

\begin{figure}[]
	\centering
	\begin{subfigure}[b]{0.45\columnwidth}
		\centering
		\includegraphics[width=\textwidth]{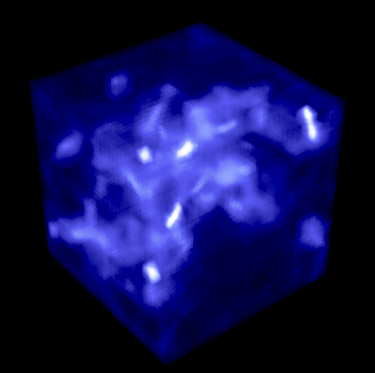}
	\end{subfigure}%
	\hfill%
	\begin{subfigure}[b]{0.45\columnwidth}
		\centering
		\includegraphics[width=\textwidth]{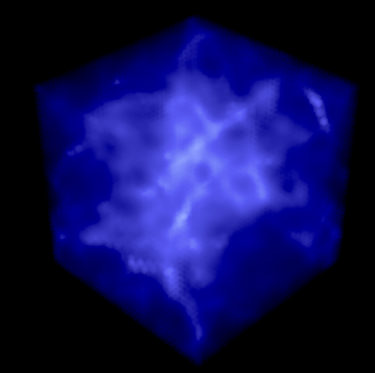}
	\end{subfigure}%
	\\%
	\begin{subfigure}[b]{0.45\columnwidth}
		\centering
		\includegraphics[width=\textwidth]{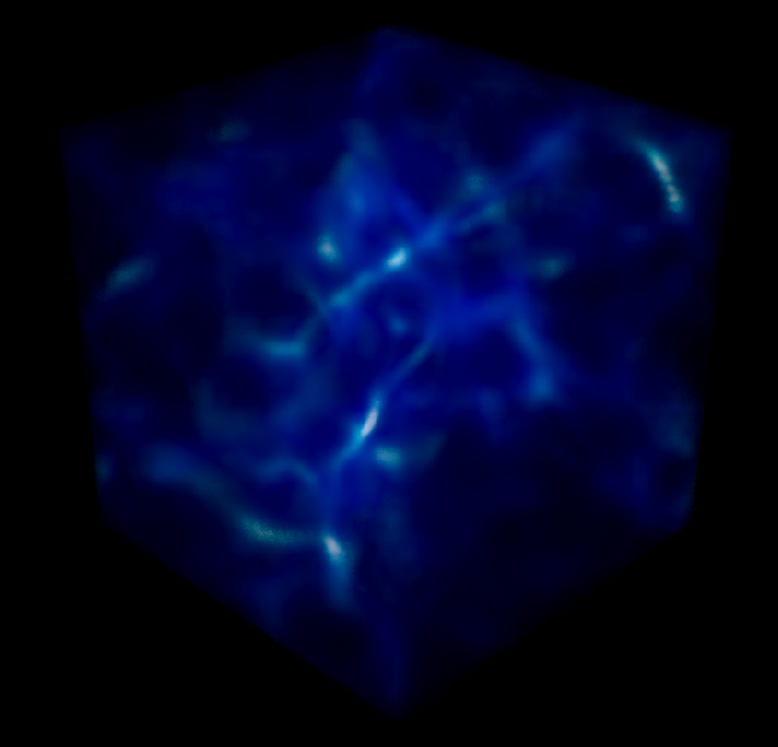}
	\end{subfigure}%
	\subfigsCaption{Enzo Tiny Cosmology rendered with our Gaussians renderer(top-left), our SciVis renderer(top-right) and in Houdini\cite{ytini_amr}(bottom)}
	\label{fig:tiny_enzo_compared}
\end{figure}

We have also done a comparative rendering of loading the same multi-grid VDB files in our SciVis renderer and have results of rendering in Houdini as provided at ytini\cite{ytini_amr}. Both of our renderers use the same transfer function that was implemented to mimic the blue to white ``Emission Color Ramp'' in Houdini.  Please see in Figure~\ref{fig:tiny_enzo_compared}, where the regions of high resolution are highlighted in white, while more crisper in our results they nevertheless show similar internal structure of the dataset. The results of the SciVis rendering look more faded and smudged in comparison and results of the Houdini rendering are dependent on the color-ramp implementation, which cannot be accessed due to closed-source nature of the software.

\section{Results}

The results are evaluated based on comparison with our SciVis renderer for ground truth(described in Section 4.4) and quality of rendering with different LODs. To highlight the capabilities of this method as a unified modeling scheme, we have used illustrative examples of modeling and rendering different scientific volume dataset types like point-clouds and AMR Volumes with minimal changes to the Gaussian pipeline. But for qualitative and quantitative estimates we have used 8-bit PSNR and frames-per-second respectively. Each renderer has its own design choices and permissible parameter ranges(e.g., we apply an opacity threshold to filter out extremely low-opacity Gaussians), hence comparisons are not entirely equitable but the performance of our renderer and image quality falls within the expected quality range.

All renderings are done through an interactive viewer that supports arbitrary screen resolutions by allowing a user to dynamically adjust the rendering resolution to match the display window. Internally, each pixel is associated with an accumulation buffer that stores HDR(high dynamic range) RGB color values in \texttt{float4} format. It uses progressive rendering, where each subframe contributes one sample per pixel, and the resulting colors are accumulated over time. Depending on the frame index for each subframe, the current color sample is added to or initialized in the accumulation buffer. In an approach that is identical to the reference implementation used in NVIDIA's OptiX sample applications, the final image displayed in the viewer is computed by averaging the accumulated color values over the number of subframes and converting the result to an 8-bit \texttt{uchar4} format for display. 

\begin{figure}[tb]
	\centering
	
	\begin{subfigure}{0.48\columnwidth}
		\includegraphics[width=\linewidth]{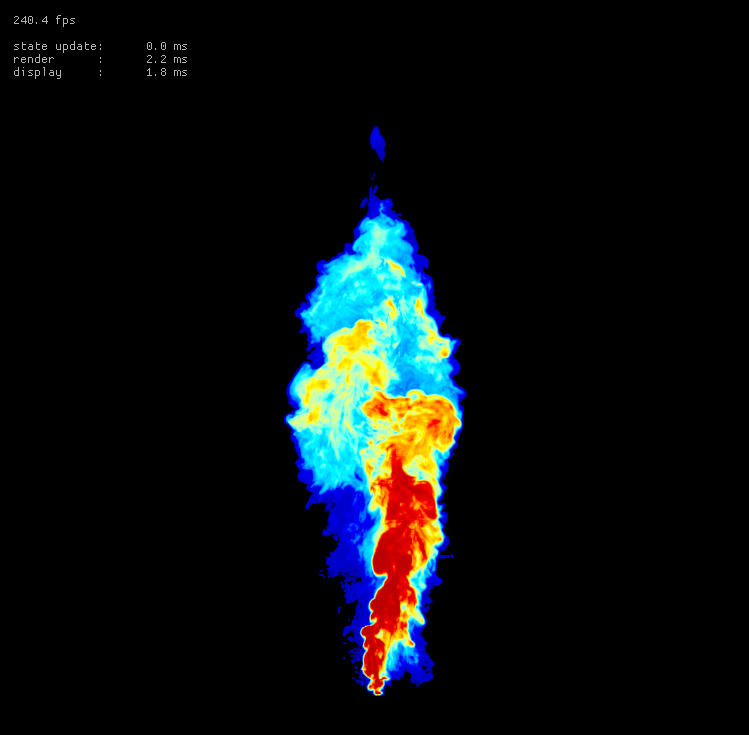}
	\end{subfigure}
	\hfill
	\begin{subfigure}{0.48\columnwidth}
		\includegraphics[width=\linewidth]{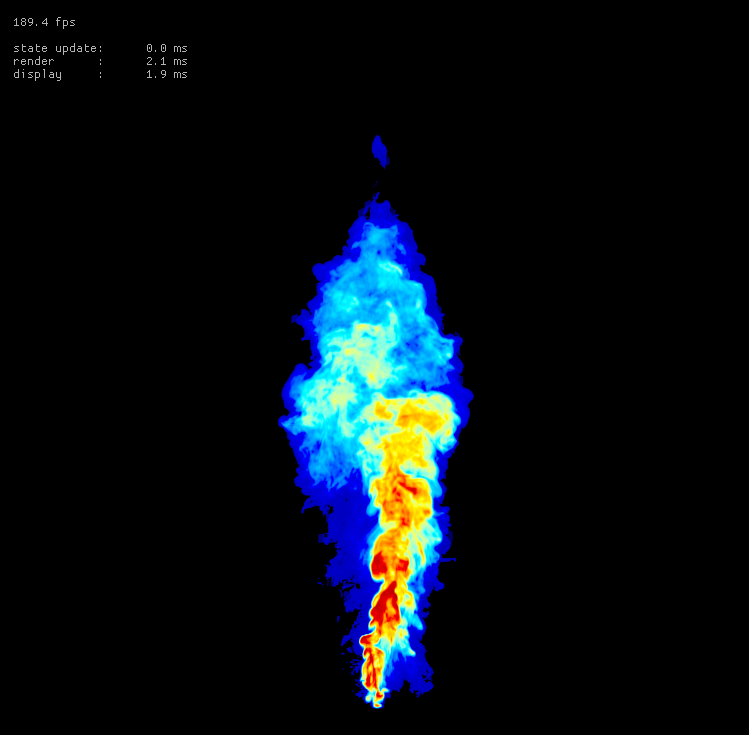}
	\end{subfigure}
	
	\begin{subfigure}{0.48\columnwidth}
		\includegraphics[width=\linewidth]{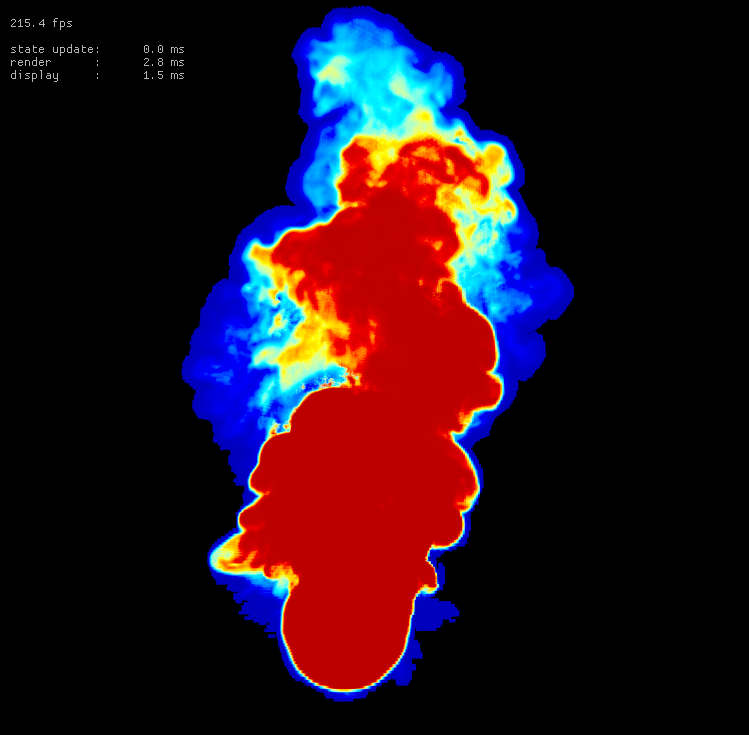}
	\end{subfigure}
	\hfill
	\begin{subfigure}{0.48\columnwidth}
		\includegraphics[width=\linewidth]{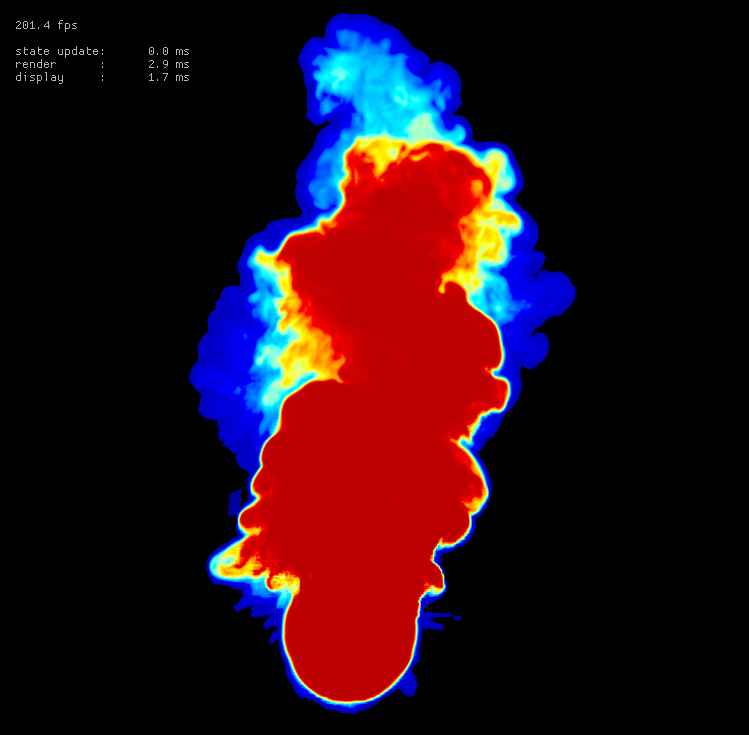}
	\end{subfigure}
	
	\begin{subfigure}{0.48\columnwidth}
		\includegraphics[width=\linewidth]{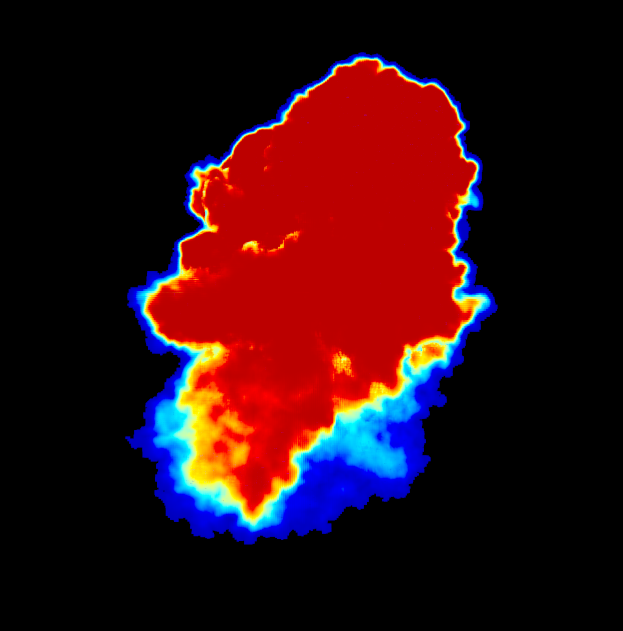}
	\end{subfigure}
	\hfill
	\begin{subfigure}{0.48\columnwidth}
		\includegraphics[width=\linewidth]{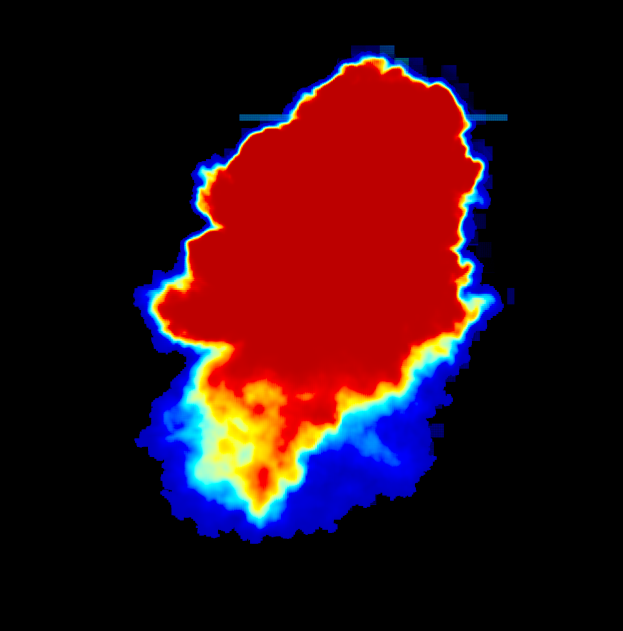}
	\end{subfigure}
	
	\caption{Comparison of the Gaussian renderer(left) vs the SciVis renderer(right) for OpenVDB smoke2, smoke, and explosion datasets.}
	\label{fig:openvdb_smoke_explosion}
\end{figure}

Figure~\ref{fig:openvdb_smoke_explosion} along with Figure~\ref{fig:openvdb_fire_bunny} presents a side-by-side comparison of five OpenVDB datasets: smoke2, smoke, explosion and fire and bunny\_cloud respectively, rendered using  our Gaussians renderer(left) and our SciVis renderer(right). The results are rendered with fine Gaussian settings for Dense Nodes($2^3$ voxels block) and 'smart grouping' strategy of Non-dense Nodes. 

\subsection{Visual comparison} A side-by-side visual comparison highlights the limitations and artifacts of our method. In the Gaussian rendering, optical depth tends to saturate slightly faster than in the SciVis renderer. This effect is caused by the inherently different shape and distribution of values: voxels assign a constant scalar value throughout their extent, whereas Gaussians concentrate values around their mean. As a result, depending on the viewing direction, some regions exhibit slightly lower optical depth than the ground truth, while others appear denser. Strategies to mitigate these effects are discussed in the Conclusion and Future Work section.

On the positive side, our approach provides full volumetric coverage of the dataset, in contrast to surface-based Gaussian representations that are ``hollow'' inside. A zoomed-in rendering of the \texttt{smoke2} dataset in Figure~\ref{fig:zoom_smoke2} illustrates this advantage particularly well.

\subsection{FPS, PSNR and Number of Gaussian primitives}

For quantitative evaluation, particularly for computing PSNR between rendered outputs, all images were captured at a fixed resolution to ensure fair and consistent comparison. Since the interactive viewer does not perform tone mapping or color correction, the 8-bit output represents a linearly scaled version of the accumulated radiance. Although this approach does not retain the full floating-point precision of the accumulation buffer, it still provides a consistent basis for comparing renderings across different configurations. All PSNR values reported are computed between these uniformly rendered and captured images, each accumulated over a fixed number of subframes.

To evaluate the reconstruction quality of our Gaussian modeling, we focus on varying the LOD across three representative volumetric datasets for both dense and non-dense leaf Nodes. Each dataset is rendered under multiple LOD settings, and the results are compared using PSNR and FPS as the main metrics. The resulting tables of evaluation are listed below:

\textbf{First evaluation(Table~\ref{tab:lod-dense}):} The non-dense regions are kept constant using the adaptive \textit{smart grouping} strategy, while the dense regions are varied across different block grouping levels: $2\times2\times2$, $4\times4\times4$, and $8\times8\times8$. This allows us to isolate the impact of block granularity in dense regions on visual fidelity.
\begin{table}[h]
	\scriptsize
	\centering
	\begin{tabu}{rccc}
		\toprule
		& explosion & smoke2 & smoke \\
		\midrule
		
		\multicolumn{4}{l}{\textbf{2×2×2 dense leaf nodes group}} \\
		PSNR & 24.02 & 24.07 & 24.11 \\
		FPS & 125.4 & 266 & 235.5 \\
		\# Gaussians & 1.5M & 2.6M & 890K \\
		
		\addlinespace
		
		\multicolumn{4}{l}{\textbf{4×4×4 dense leaf nodes group}} \\
		PSNR & 28.22 & 23.96 & 23.41 \\
		FPS & 226.6 & 333 & 324 \\
		\# Gaussians & 1.3M & 2.5M & 871K \\
		
		\addlinespace
		
		\multicolumn{4}{l}{\textbf{8×8×8 dense leaf nodes group}} \\
		PSNR & 21.87 & 25.00 & 25.70 \\
		FPS & 248 & 342 & 327 \\
		\# Gaussians & 1.2M & 2.4M & 868K \\
		
		\bottomrule
	\end{tabu}
    \caption{
		PSNR, FPS and number of Gaussians evaluation across different dense leaf node grouping strategies($2^3$, $4^3$ and $8^3$, with non-dense nodes fitted using the `smart grouping' method.}
        \label{tab:lod-dense}
\end{table}

\begin{figure}[h]
	\centering
	
	\begin{subfigure}{0.30\columnwidth}
		\includegraphics[width=\linewidth]{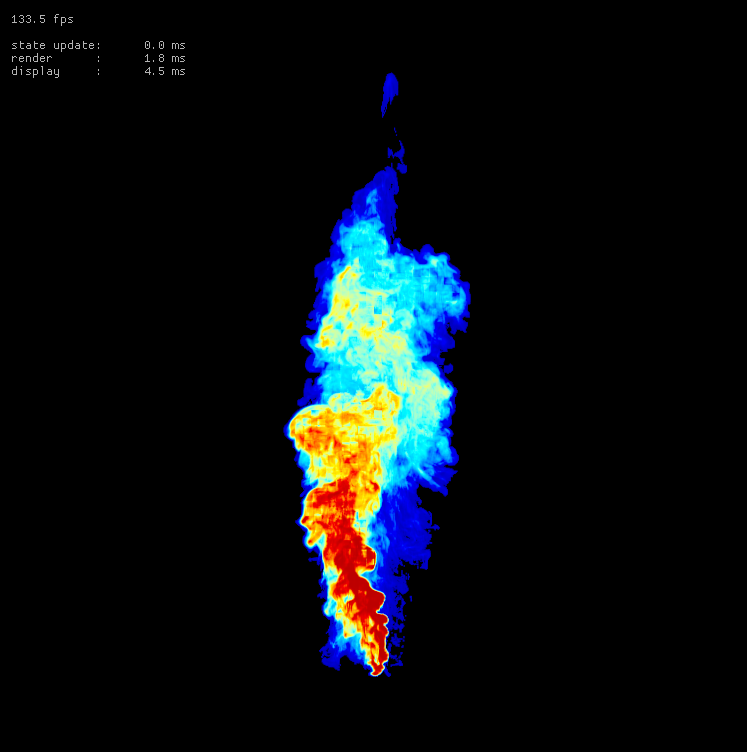}

	\end{subfigure}
	\hfill
	\begin{subfigure}{0.30\columnwidth}
		\includegraphics[width=\linewidth]{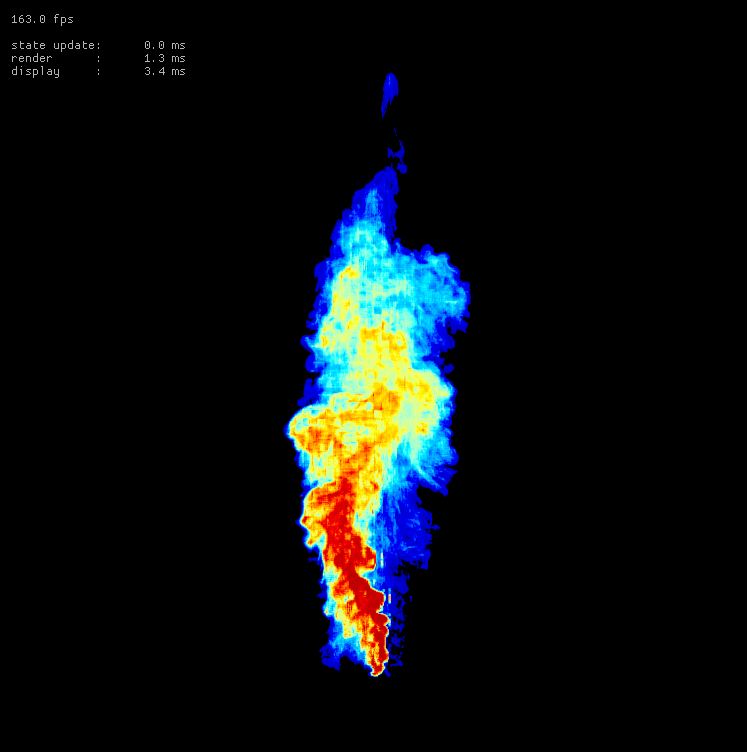}

	\end{subfigure}
	\hfill
	\begin{subfigure}{0.30\columnwidth}
		\includegraphics[width=\linewidth]{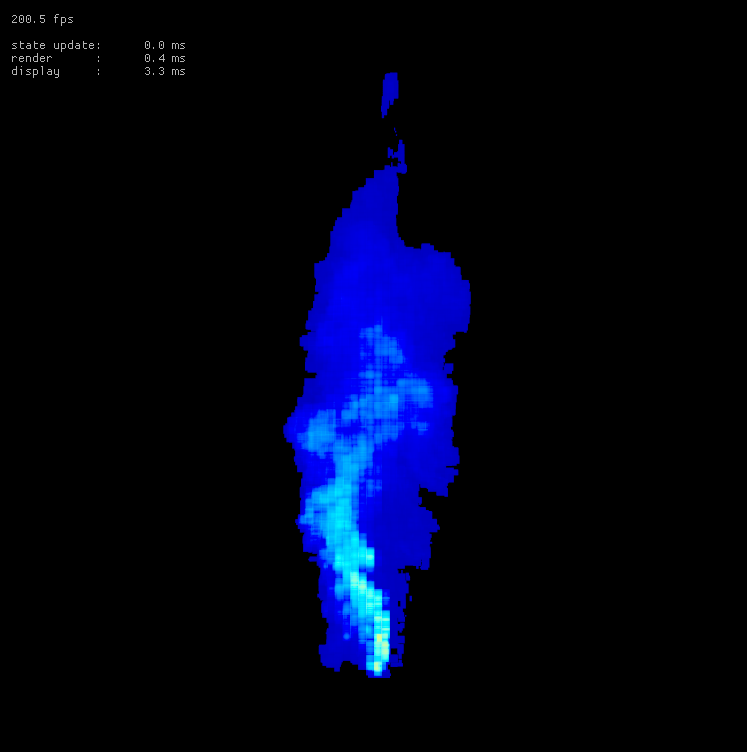}

	\end{subfigure}
    
    \begin{subfigure}{0.30\columnwidth}
		\includegraphics[width=\linewidth]{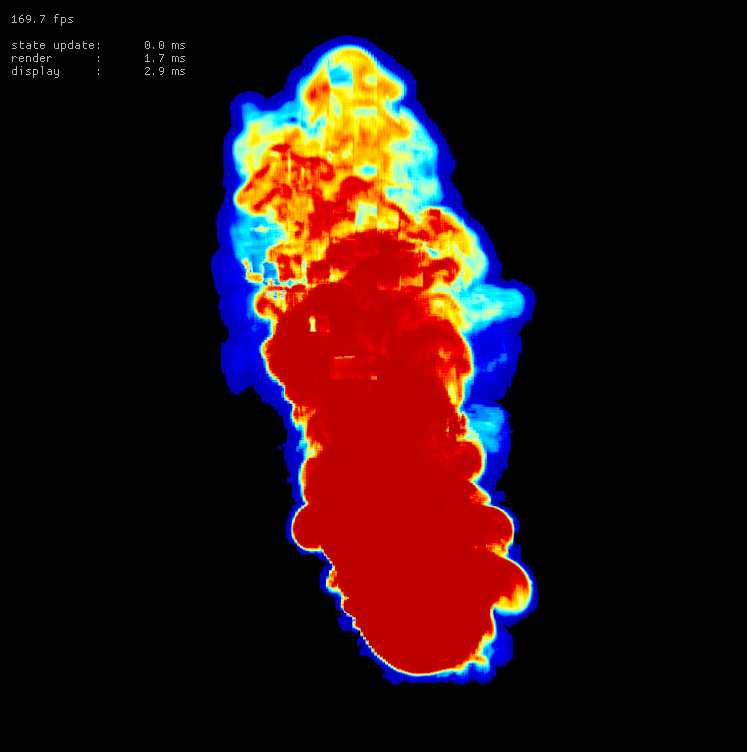}

	\end{subfigure}
	\hfill
	\begin{subfigure}{0.30\columnwidth}
		\includegraphics[width=\linewidth]{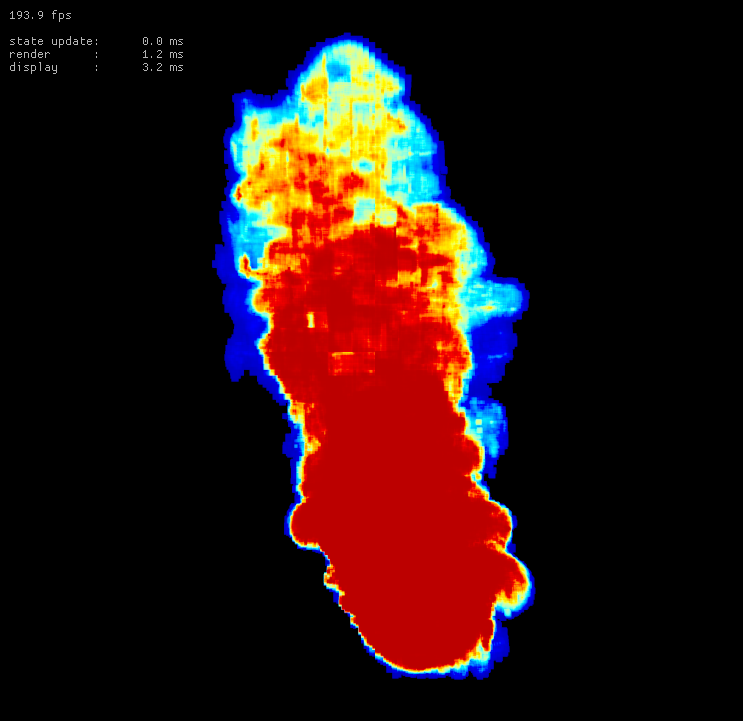}

	\end{subfigure}
	\hfill
	\begin{subfigure}{0.30\columnwidth}
		\includegraphics[width=\linewidth]{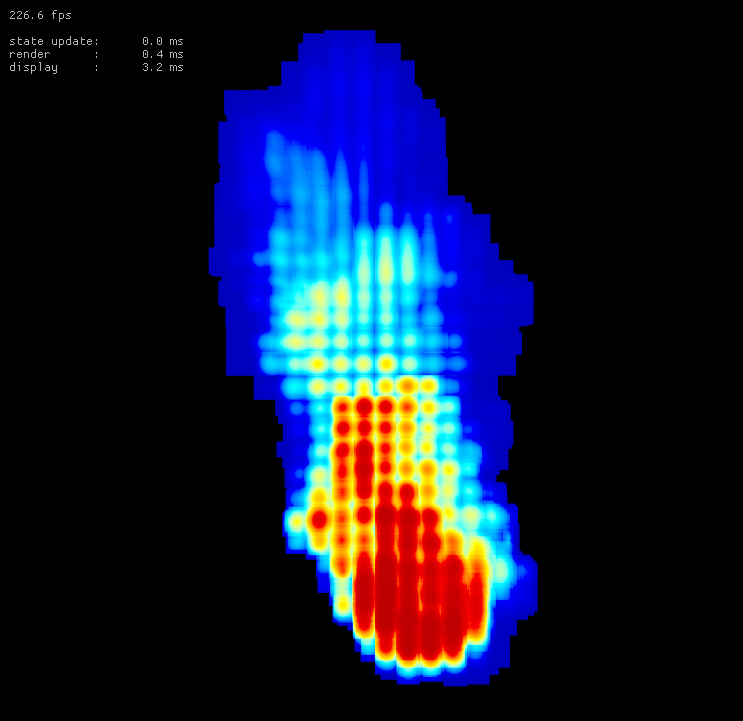}

	\end{subfigure}
	
	\begin{subfigure}{0.30\columnwidth}
		\includegraphics[width=\linewidth]{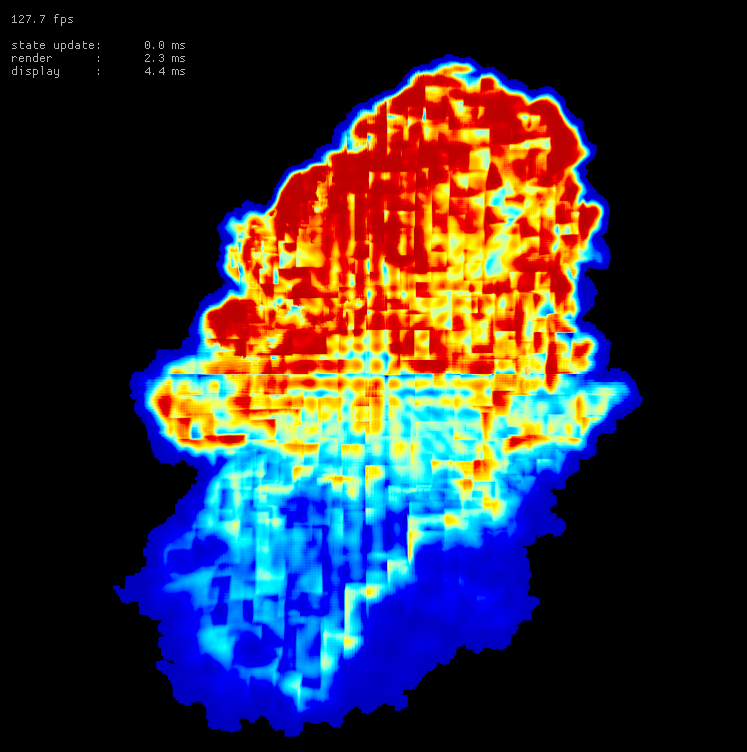}

	\end{subfigure}
	\hfill
	\begin{subfigure}{0.30\columnwidth}
		\includegraphics[width=\linewidth]{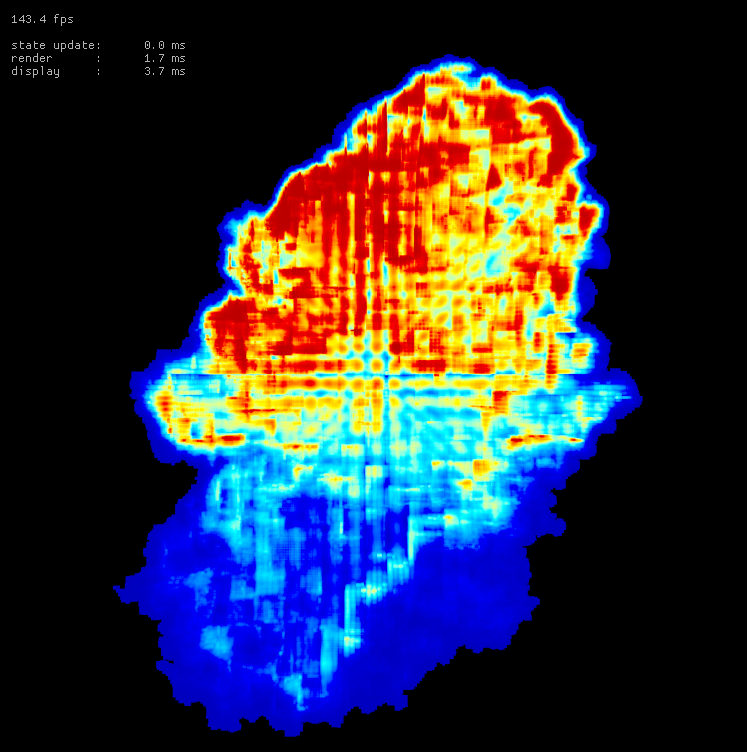}

	\end{subfigure}
	\hfill
	\begin{subfigure}{0.30\columnwidth}
		\includegraphics[width=\linewidth]{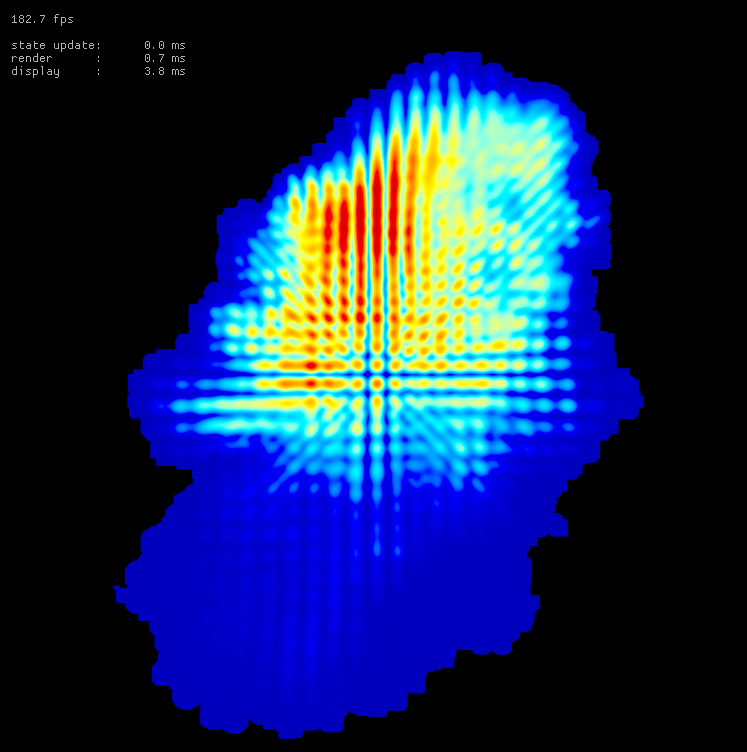}

	\end{subfigure}
	
	\caption{%
	Visual comparison of three volumetric datasets(top to bottom: smoke2, smoke and explosion) with three non-dense Gaussian grouping strategies(from left to right): \textit{smart grouping}, \textit{strict blocks + single voxels}, and \textit{single Gaussian per leaf} while keeping dense nodes at the coarsest leaf-node resolution).
	}
	\label{fig:non_dense_comparison}
\end{figure}

\textbf{Second evaluation(Table~\ref{tab:lod-nondense}):} We fix the dense node approximation at the coarsest possible level(entire $8^3$ leaf nodes) and progressively reduce the fidelity of the non-dense strategy, from smart grouping, to strict $2^3$ blocks with fallback to single voxels, and finally to a single Gaussian per sparse leaf. Figure~\ref{fig:non_dense_comparison} illustrates the visual differences across three non-dense Gaussian grouping strategies applied to the smoke, smoke2, and explosion datasets. Because the dense regions are fixed to a very coarse approximation, any difference in visual fidelity will largely stem from how the sparse(non-dense) regions are grouped, which mostly occupy the periphery of a model. This isolates the contribution of the non-dense grouping strategy and illustrates how aggressive simplification in sparse regions affects overall quality. 

\begin{table}[]
	\scriptsize
	\centering
	\begin{tabu}{rccc}
		\toprule
		& explosion & smoke2 & smoke \\
		\midrule
		
		\multicolumn{4}{l}{\textbf{Non-dense 2×2×2(smart grouping)}} \\
		PSNR & 21.87 & 25.00 & 25.70 \\
		FPS & 248 & 342 & 327 \\
		\# Gaussians & 1.2M & 2.4M & 868K \\
		
		\addlinespace
		
		\multicolumn{4}{l}{\textbf{Non-dense 2×2×2(strict blocks + single voxels)}} \\
		PSNR & 21.30 & 24.75 & 22.43 \\
		FPS & 144 & 163 & 194 \\
		\# Gaussians & 329K & 749K & 153K \\
		
		\addlinespace
		
		\multicolumn{4}{l}{\textbf{Non-dense(single biggest Gaussian per leaf)}} \\
		PSNR & 18.90 & 23.02 & 20.87 \\
		FPS & 183 & 200 & 227 \\
		\# Gaussians & 10K & 10.1K & 2.5K \\
		
		\bottomrule
	\end{tabu}
    \caption{%
		PSNR, FPS and number of Gaussians evaluation across different sparse leaf Node grouping strategies, with dense nodes fitted using the coarsest method.
	}
    \label{tab:lod-nondense}
\end{table}

\begin{figure}[h]
	\centering
	
	\begin{subfigure}{0.30\columnwidth}
		\includegraphics[width=\linewidth]{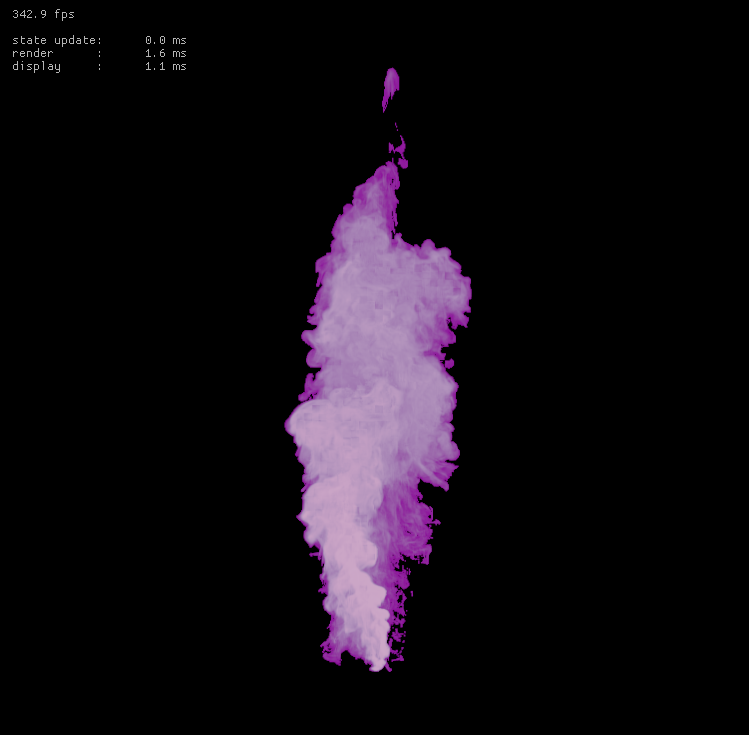}

	\end{subfigure}
	\hfill
	\begin{subfigure}{0.30\columnwidth}
		\includegraphics[width=\linewidth]{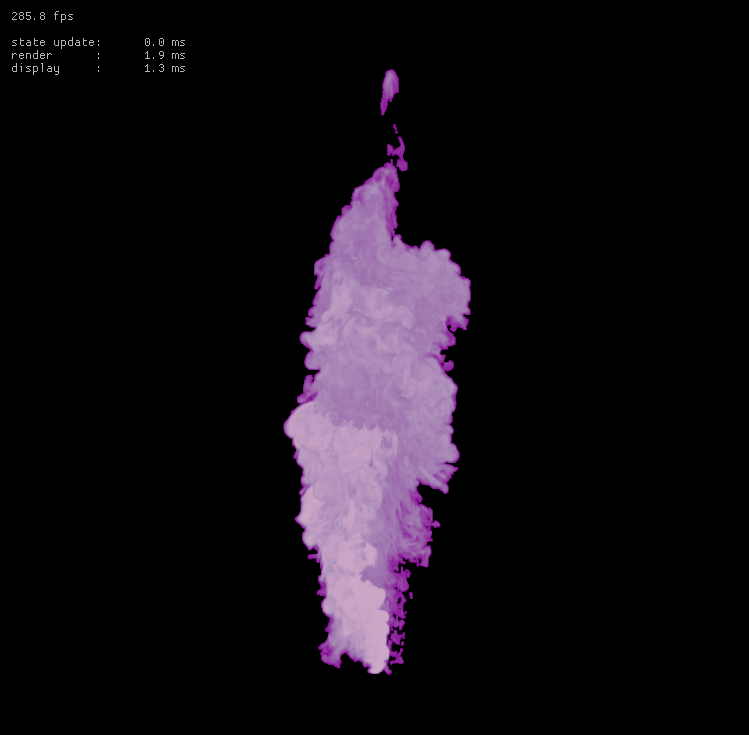}

	\end{subfigure}
	\hfill
	\begin{subfigure}{0.30\columnwidth}
		\includegraphics[width=\linewidth]{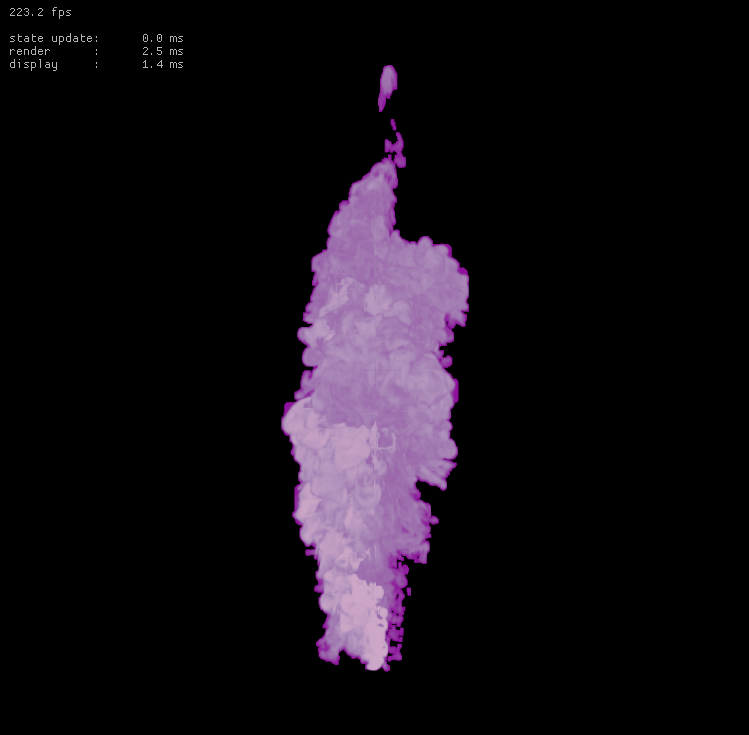}

	\end{subfigure}

	\begin{subfigure}{0.30\columnwidth}
		\includegraphics[width=\linewidth]{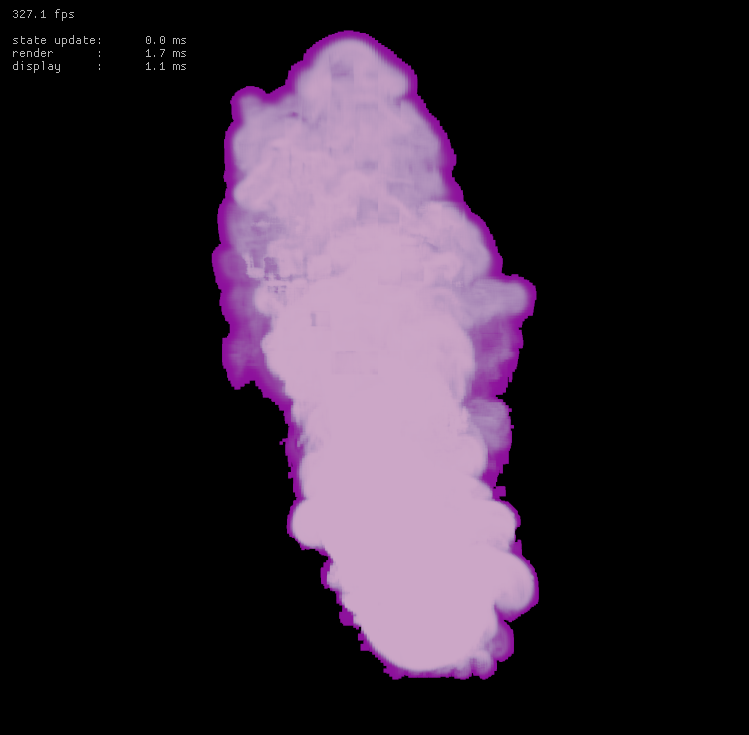}

	\end{subfigure}
	\hfill
	\begin{subfigure}{0.30\columnwidth}
		\includegraphics[width=\linewidth]{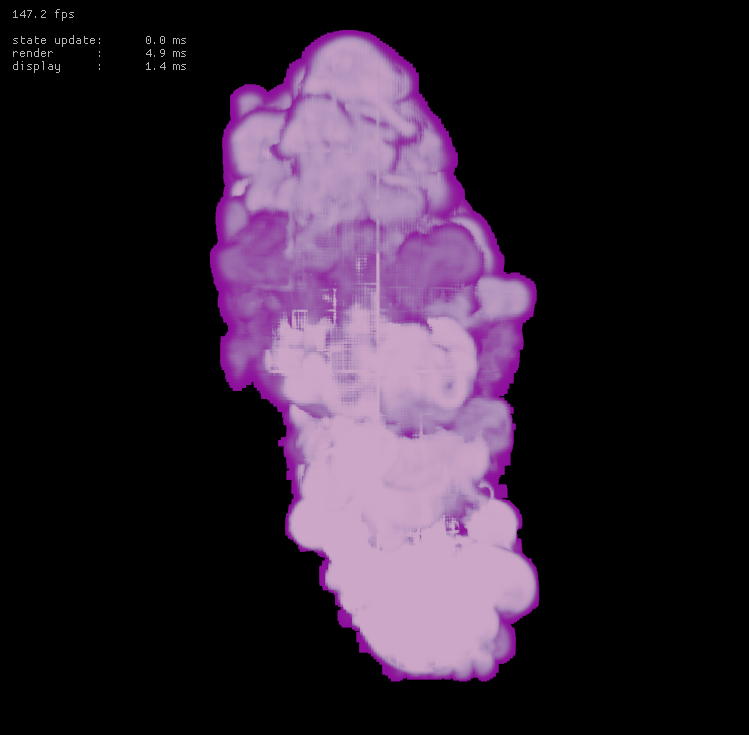}

	\end{subfigure}
	\hfill
	\begin{subfigure}{0.30\columnwidth}
		\includegraphics[width=\linewidth]{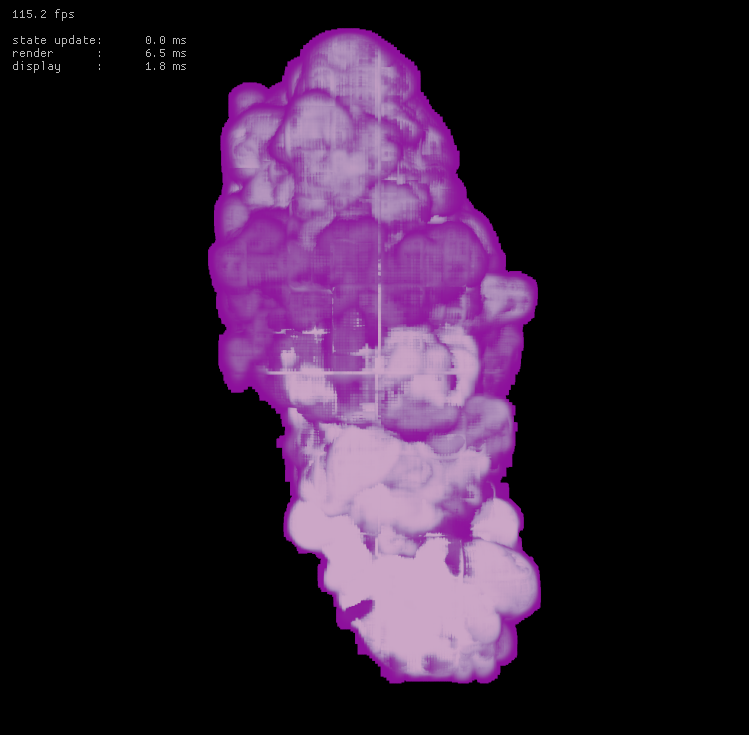}

	\end{subfigure}

    \begin{subfigure}{0.30\columnwidth}
		\includegraphics[width=\linewidth]{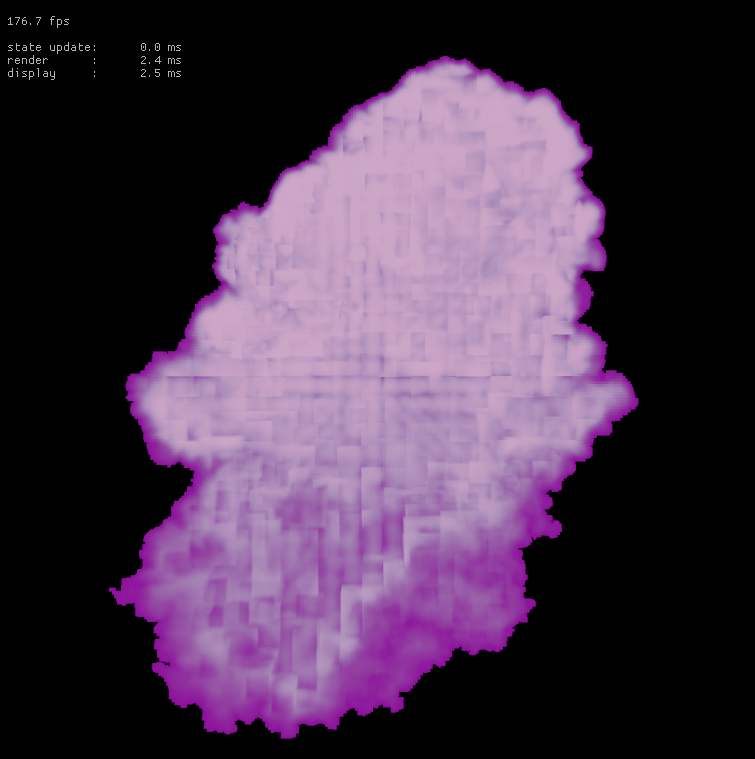}

	\end{subfigure}
	\hfill
	\begin{subfigure}{0.30\columnwidth}
		\includegraphics[width=\linewidth]{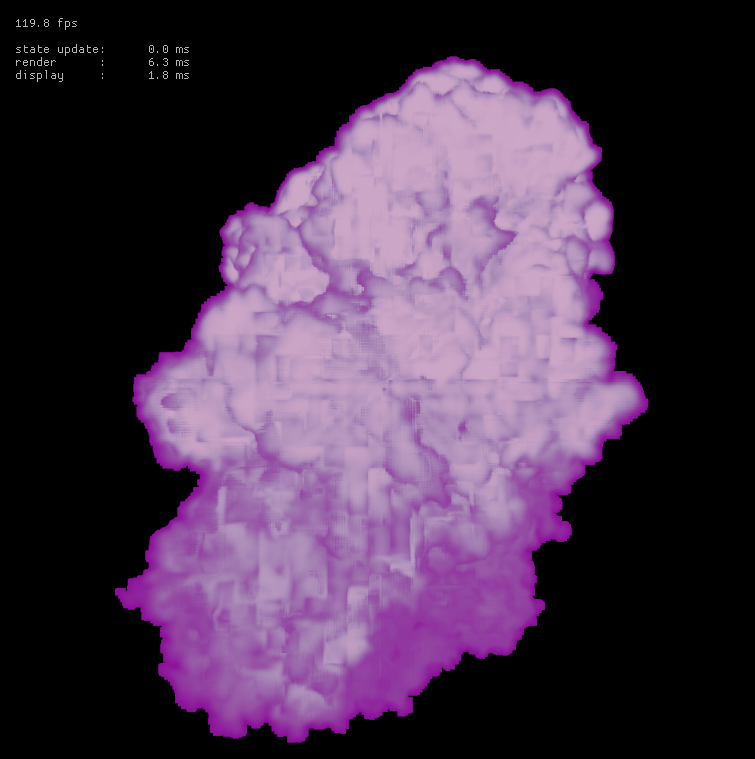}

	\end{subfigure}
	\hfill
	\begin{subfigure}{0.30\columnwidth}
		\includegraphics[width=\linewidth]{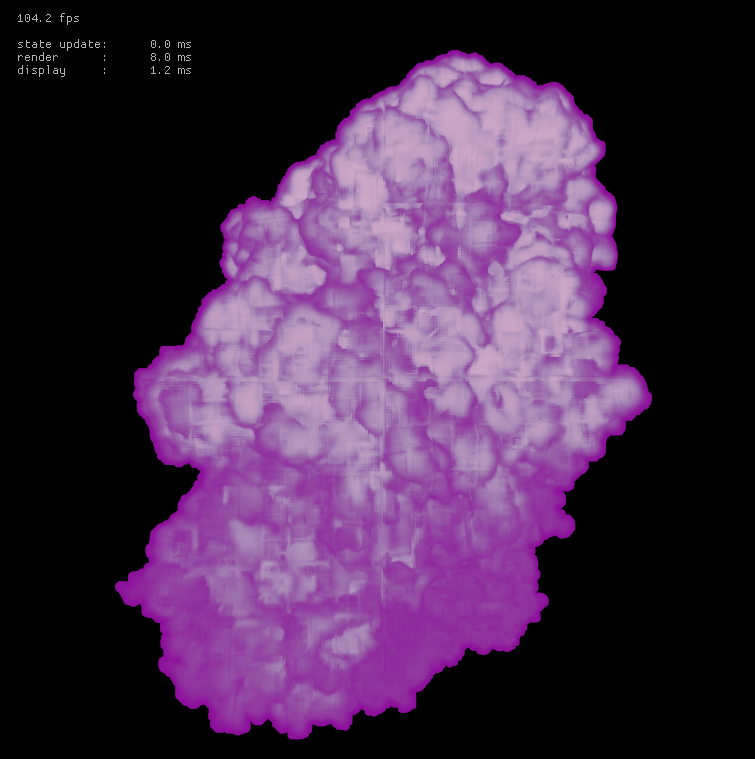}

	\end{subfigure}
	
	\caption{%
		Comparison of different Gaussian extent multipliers(from left to right: \(\sigma = 1, 2, 3\)) applied to the smoke2, smoke and explosion datasets(top to bottom). 
	}

	\label{fig:sigma_comparison}
\end{figure}

 \textbf{Third evaluation(Table~\ref{tab:sigma-comparison}):} It explores the effect of varying the Gaussian bounding box extent by modifying the standard deviation multiplier(\(\sigma\)) used to define the AABB. This directly impacts the tightness of spatial acceleration structures and indirectly affects rendering performance and accuracy. As shown in Figure~\ref{fig:sigma_comparison}, varying the Gaussian extent multiplier(\(\sigma\)) significantly affects the visual representation of the three datasets.

\begin{table}[h]
	\scriptsize
	\centering
	\begin{tabu}{rccc}
		\toprule
		& explosion & smoke2 & smoke \\
		\midrule
		
		\multicolumn{4}{l}{\textbf{8×8×8 dense leaf nodes(sigma = 1)}} \\
		PSNR & 21.87 & 25.00 & 25.70 \\
		FPS & 248 & 342 & 327 \\
		\# Gaussians & 1.2M & 2.4M & 868K \\
		
		\addlinespace
		
		\multicolumn{4}{l}{\textbf{8×8×8 dense leaf nodes(sigma = 2)}} \\
		PSNR & 20.80 & 24.69 & 22.00 \\
		FPS & 210 & 316 & 285 \\
		\# Gaussians & 1.2M & 2.4M & 868K \\
		
		\addlinespace
		
		\multicolumn{4}{l}{\textbf{8×8×8 dense leaf nodes(sigma = 3)}} \\
		PSNR & 18.01 & 24.21 & 19.23 \\
		FPS & 176 & 279 & 254 \\
		\# Gaussians & 1.2M & 2.4M & 868K \\
		
		\bottomrule
	\end{tabu}
    \caption{%
		PSNR, FPS and number of Gaussians evaluation for coarse leaf Node group fitting with different sigma values.
	}
    \label{tab:sigma-comparison}
\end{table}

\begin{figure}[]
	\centering
	\includegraphics[width=\columnwidth]{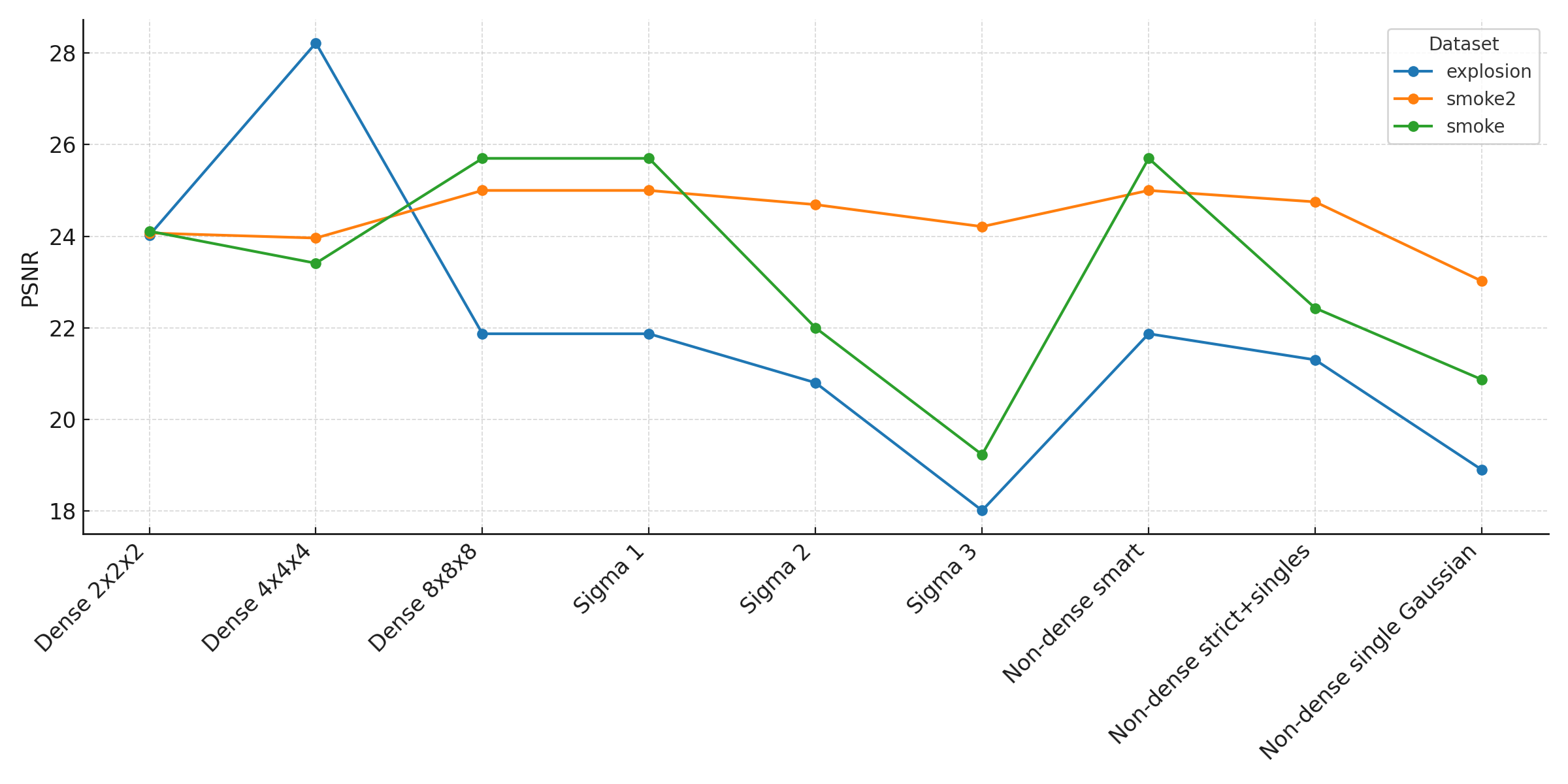}
	\caption{%
		Peak signal-to-noise ratio(PSNR) across different Gaussian grouping and AABB extent strategies for three volumetric datasets.
	}
	\label{fig:psnr_comparison}
\end{figure}

\begin{figure}[tb]
	\centering
	\includegraphics[width=\columnwidth]{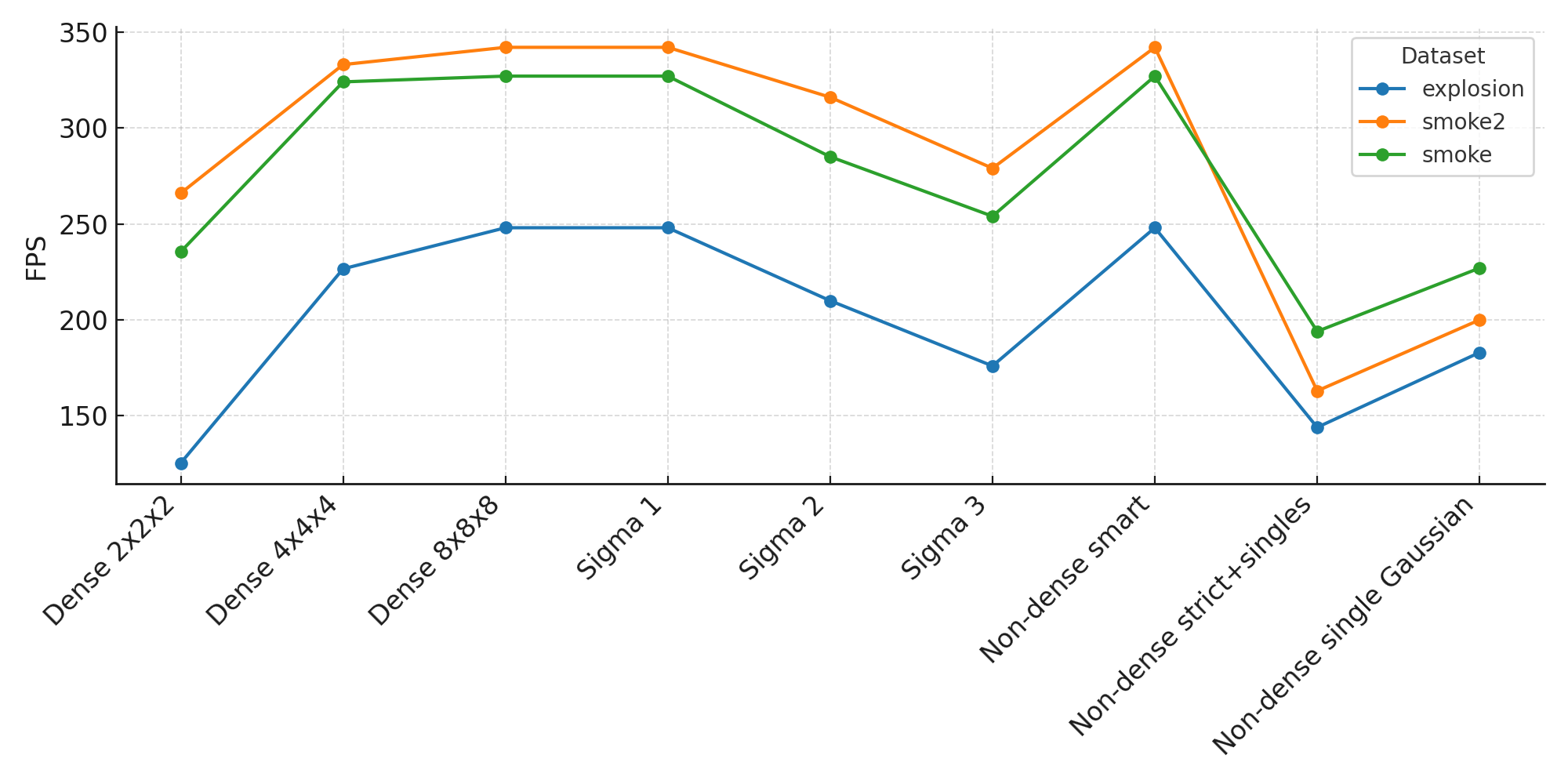}
	\caption{%
		Rendering performance(frames per second) for various Gaussian grouping and AABB extent strategies evaluated on explosion, smoke2, and smoke datasets.
	}
	\label{fig:fps_comparison}
\end{figure}

\textbf{LOD Trade-offs for PSNR and FPS:} 
Figures~\ref{fig:psnr_comparison} and~\ref{fig:fps_comparison} present the quantitative evaluation of different Gaussian grouping strategies across the three volumetric datasets. Along the x-axis in both the figures are listed various LODs achieved by incorporating different grouping strategies. The dense strategies on the left have constant non-dense grouping using ``Smart Grouping'' scheme and non-dense strategies on the right, have dense grouping at the coarsest LOD. Figure~\ref{fig:psnr_comparison} shows the peak signal-to-noise ratio(PSNR), highlighting how finer grouping preserves more detail for 2 out of 3 datasets, particularly in dense regions. In contrast, Figure~\ref{fig:fps_comparison} reports the rendering performance in terms of the FPS, illustrating the trade-off between visual quality and computational efficiency. As expected, coarser groupings yield higher frame rates but lower PSNR, while finer groupings improve reconstruction fidelity although still giving interactive frame rates. 

\textbf{Number of primitives compression:} 
Finally, for comparison of raw number of Gaussians to number of voxels we have chosen 5 LOD levels:
\begin{itemize}
	\item LOD-1: Dense $2^3$ + Non-dense Smart Grouping
	\item LOD-2: Dense $4^3$ + Non-dense Smart Grouping
	\item LOD-3: Dense $8^3$ + Non-dense Smart Grouping
	\item LOD-4: Dense $8^3$ + Non-dense Strict Blocks + Single Voxels
	\item LOD-5: Dense $8^3$ + Non-dense Single Gaussian Per Leaf
\end{itemize}

Table~\ref{tab:lod_reduction} summarizes the number of Gaussians generated for each data set in the five distinct LOD configurations listed above, ranging from fine-grained to highly compact representations. 
\begin{table}[h]
    \centering
    \begin{tabu}{lccc}
        \toprule
        \textbf{LOD} & \textbf{explosion} & \textbf{smoke2} & \textbf{smoke} \\
        \midrule
        Voxels & 12.58M (100\%) & 20.74M (100\%) & 2.76M (100\%) \\
        LOD-1  & 1.50M (11.9\%) & 2.60M (12.5\%) & 0.89M (32.2\%) \\
        LOD-2  & 1.30M (10.3\%) & 2.50M (12.1\%) & 0.87M (31.6\%) \\
        LOD-3  & 1.20M (9.5\%)  & 2.40M (11.6\%) & 0.87M (31.4\%) \\
        LOD-4  & 0.33M (2.6\%)  & 0.75M (3.6\%)  & 0.15M (5.5\%)  \\
        LOD-5  & 0.01M (0.08\%) & 0.01M (0.05\%) & 0.0025M (0.09\%) \\
        \bottomrule
    \end{tabu}
    \caption{Primitive count reduction across five LODs compared to voxel counts for three datasets. Percentages indicate fraction relative to the voxel baseline.}
    \label{tab:lod_reduction}
\end{table}
Across all three datasets, Gaussian primitives reduce dramatically in number compared to voxels, with reductions spanning up to three orders of magnitude at higher LODs. At LOD-1, the explosion and smoke2 datasets retain only about 12\% of their voxel count, while smoke remains relatively denser at 32\%. From LOD-2 to LOD-3, explosion and smoke2 stabilize in the 10--12\% range, whereas smoke consistently retains around 31\% meaning that maximum number of Nodes in this dataset are non-dense Nodes. At LOD-4, the primitive count drops sharply, with explosion and smoke2 reduced to below 4\% and smoke to around 5.5\% which is counter-intuitive because `smart grouping' strategy should ideally produce fewer than its strict counterpart. An explanation of this is far fewer Gaussians making the cut for opacity threshold checks in the case of the latter. Finally, LOD-5 achieves the strongest compression, with all datasets reduced to less than 0.1\% of their original voxel counts, representing reductions of nearly three orders of magnitude. This demonstrates the efficiency of the Gaussian representation in significantly lowering primitive counts while maintaining volumetric coverage.

\section{Conclusion and Future Work}
\textit{Comparison with machine-learning–based optimization methods:} 
In alignment with the recent trend of creating Gaussian representations of surfaces and entire scenes, we extend this paradigm to volumetric data. The increasing interest in Gaussians as a unified primitive for graphics and vision pipelines suggests a possible future where rendering is deeply intertwined with Gaussian representations, potentially supported by specialized hardware accelerators. Matrix–matrix operations that naturally arise in Gaussian transformations, such as those for animation or morphing, could be efficiently mapped to such accelerators, further motivating this shift. Unlike machine-learning–based Gaussian optimization methods, which typically rely on differentiable rasterization and are trained directly on image data, our approach leverages volumetric inputs directly. While differentiable rasterization enables high-quality reconstructions from training viewpoints, it does not guarantee volumetric consistency\cite{celarek2025gaussiansplatting}. Unless the dataset contains close-up observations across all regions, the optimized Gaussians capture only visible surfaces. As a result, zooming into such reconstructions often reveals hollow interiors rather than continuous volumetric structure. In contrast, our method generates Gaussians from volumetric data itself, ensuring complete coverage and avoiding this limitation.

\textit{Dense versus non-dense nodes in OpenVDB:} 
Our approach to Gaussian generation is guided by the spatial distribution of data within OpenVDB volumes, which consist of both dense and non-dense Nodes. Since these node types differ significantly in data distribution and effective center of mass, they must be handled differently in the Gaussian generation process. OpenVDB provides well-suited spatial statistics for Gaussian initialization, such as block centroids or leaf-node centroids for positioning, allowing us to produce effective Gaussian approximations without end-to-end machine learning. The performance of our method is strongly influenced by the ratio of dense to non-dense nodes in a given dataset. Depending on this distribution, it is possible to achieve very good results even under the coarsest settings. This is a general challenge in data fitting: the quality of approximation is inherently tied to both the underlying structure of the data and the diversity of values it contains.

\textit{Variance-aware strategies:} 
One promising direction for improving quality of the fitting is to employ variance-aware strategies, where nodes with higher internal variance are treated differently from more homogeneous regions. Such methods would allow the Gaussian representation to adapt more flexibly to local data complexity. We experimented with variance-based approaches\cite{Sharma2025GaussianParticleApproximation}. However, the datasets available for our initial study exhibited relatively low variance within a leaf Node, which limited their effectiveness and it did not produce improvements sufficient to justify its inclusion in the current pipeline at the cost of calculating proper covariance or a voxel-local gradient between 512 voxels to identify regions of high variance. If the domain spans multiple leaf Nodes, such as through greedy grouping, the analysis of variance becomes significantly more complex. Nevertheless, we consider variance-aware modeling to be a promising direction for future refinement. However, pursuing this would require a deep investigation into a parallel topic, while potentially valuable, it constitutes a substantial research area.

\textit{OptiX AABB limitation:} 
Our renderer is constrained by OptiX’s reliance on axis-aligned bounding boxes (AABBs). A more faithful enclosure of anisotropic Gaussians would ideally use oriented bounding boxes (OBBs) or other tighter volumetric bounds, like the bounding geometries in the work of Moenne-Loccoz et al.\cite{MoenneLoccoz2024}. The use of AABBs can lead to inefficiencies in intersection handling and sampling, as the bounding volumes are not always well aligned with the Gaussian covariance structure. Moreover, enforcing non-overlapping Gaussians for efficient clustering often introduces gaps or a loss of fidelity. This highlights a fundamental mismatch: unlike rectilinear voxel grids, Gaussians cannot uniformly tile space without trade-offs in coverage or compactness. Addressing these limitations will require either improved acceleration structures or hybrid approaches that combine the strengths of Gaussian and voxel representations.

\textit{Opacity over and under-saturation:}
One of the key limitations we observe is related to opacity saturation artifacts, where Gaussians can lead to over- or under-saturation of opacity with optical depth calculations in areas of aligning high Gaussian densities such as centers or low such as edges respectively. Addressing this issue requires moving beyond strictly non-overlapping primitives toward richer Gaussian mixture models in which overlapping Gaussians accumulate more faithfully to the target values. Achieving such overlap requires a deeper global understanding of the data rather than relying solely on local leaf Node statistics. While our current method performs well for the datasets considered, applications demanding very high fidelity will likely require true mixture models. Constructing these models would involve flattening the OpenVDB tree and applying clustering algorithms across the entire dataset to capture global correlations. However, this comes at the cost of one of OpenVDB’s main benefits: its hierarchical sparsity and raises scalability challenges. Nonetheless, such an approach has the potential to eliminate opacity saturation related artifacts by ensuring that overlapping Gaussians accumulate to the desired volumetric values.

\textit{Hybrid approach with Machine-learned mixture models:}
In relation to the paragraphs above, optimizing Gaussian mixtures against actual data values by utilizing an error metric as the basis of optimization such as the L2 norm between the rendered optical depth and the ground-truth distribution, along with deterministic Gaussian initialization can produce very quick convergence to the ground truth and alleviate some of the problems of the current method. We are actively investigating this direction and expect that hybrid approaches, combining OpenVDB-driven initialization with machine-learning–based refinement, will enable higher fidelity results and more robust generalization to diverse volumetric datasets.

Finally, despite the above limitations imposed by non-overlapping clustering and axis-aligned bounding constraints, our method maintains the structural integrity of the original volumetric data. Unlike surface-based Gaussian splatting approaches, which often result in hollow or distorted representations when viewed up close, our volumetric Gaussians encode actual volume. As a result, users can interactively zoom into the rendering at any scale and still observe a structure that closely resembles voxel-based volume rendering(see Figure~\ref{fig:zoom_smoke2}). This fidelity supports detailed spatial exploration while significantly reducing the number of primitives. Our results position volumetric Gaussians as a compelling middle ground between dense voxel grids and lightweight surface abstractions. We are further exploring more expressive models that can address existing limitations in coverage, overlapping behavior and clustering flexibility.

\bibliographystyle{abbrv-doi-hyperref}
\bibliography{main}

\begin{thebibliography}{10}

\bibitem{bailey2015distributing}
D.~Bailey, H.~Biddle, N.~Avramoussis, and M.~Warner.
\newblock Distributing liquids using openvdb.
\newblock In {\em ACM SIGGRAPH 2015 Talks}, 2015. \href{https://doi.org/10.1145/2775280.2792544}
{doi: {{%
10\hspace{.1pt}\discretionary{.}{%
}{.}\hspace{.4pt}1145\discretionary{/}{%
}{/}2775280\hspace{.1pt}\discretionary{.}{%
}{.}\hspace{.4pt}2792544}}}


\bibitem{borkiewicz2017communicating}
K.~Borkiewicz, A.~Christensen, and J.~Stone.
\newblock Communicating science through visualization in an age of alternative facts.
\newblock In {\em SIGGRAPH Asia 2017 Courses}, 2017. \href{https://doi.org/10.1145/3134472.3134488}
{doi: {{%
10\hspace{.1pt}\discretionary{.}{%
}{.}\hspace{.4pt}1145\discretionary{/}{%
}{/}3134472\hspace{.1pt}\discretionary{.}{%
}{.}\hspace{.4pt}3134488}}}


\bibitem{celarek2025gaussiansplatting}
A.~Celarek, G.~Kopanas, G.~Drettakis, M.~Wimmer, and B.~Kerbl.
\newblock Does 3d gaussian splatting need accurate volumetric rendering?
\newblock {\em Computer Graphics Forum}, 44(2), 2025.
\newblock Open access; analysis of approximations in 3D Gaussian Splatting, comparing opacity-based splatting and extinction-based volumetric rendering.

\bibitem{intel_onetbb}
I.~Corporation.
\newblock Intel® oneapi threading building blocks (onetbb).
\newblock \url{https://www.intel.com/content/www/us/en/developer/tools/oneapi/onetbb.html}, 2024.

\bibitem{Dyken2025VolumeEncodingGaussians}
L.~Dyken, A.~Sewell, W.~Usher, S.~Petruzza, and S.~Kumar.
\newblock Volume encoding gaussians: Transfer‑function‑agnostic 3d gaussians for volume rendering.
\newblock {\em arXiv preprint arXiv:2504.13339}, Apr. 2025.

\bibitem{jang2006enhancing}
Y.~Jang, R.~Botchen, A.~Lauser, D.~S. Ebert, K.~P. Gaither, and T.~Ertl.
\newblock Enhancing the interactive visualization of procedurally encoded multifield data with ellipsoidal basis functions.
\newblock {\em Computer Graphics Forum}, 25(3):587--596, 2006. \href{https://doi.org/10.1111/j.1467-8659.2006.00978.x}
{doi: {{%
10\hspace{.1pt}\discretionary{.}{%
}{.}\hspace{.4pt}1111\discretionary{/}{%
}{/}j\hspace{.1pt}\discretionary{.}{%
}{.}\hspace{.4pt}1467\discretionary{%
}{-}{-}8659\hspace{.1pt}\discretionary{.}{%
}{.}\hspace{.4pt}2006\hspace{.1pt}\discretionary{.}{%
}{.}\hspace{.4pt}00978\hspace{.1pt}\discretionary{.}{%
}{.}\hspace{.4pt}x}}}


\bibitem{JubaVarshney2007GaussianRBFVolume}
D.~Juba and A.~Varshney.
\newblock Modelling and rendering large volume data with gaussian radial basis functions.
\newblock Technical Report CAR-TR-1022 / UMIACS-TR-2007-22, University of Maryland, College Park, Apr. 2007.
\newblock Technical Report.

\bibitem{kerbl3Dgaussians}
B.~Kerbl, G.~Kopanas, T.~Leimk{\"u}hler, and G.~Drettakis.
\newblock 3d gaussian splatting for real-time radiance field rendering.
\newblock {\em ACM Transactions on Graphics}, 42(4), July 2023.

\bibitem{kerbl20233dgaussiansplattingrealtime}
B.~Kerbl, G.~Kopanas, T.~Leimkühler, and G.~Drettakis.
\newblock 3d gaussian splatting for real-time radiance field rendering, 2023.

\bibitem{Lazzaro2002RBFInterpolation}
D.~Lazzaro and L.~B. Montefusco.
\newblock Radial basis functions for the multivariate interpolation of large scattered data sets.
\newblock {\em Journal of Computational and Applied Mathematics}, 140(1–2):521--536, 2002. \href{https://doi.org/10.1016/S0377-0427(01)00485-X}
{doi: {{%
10\hspace{.1pt}\discretionary{.}{%
}{.}\hspace{.4pt}1016\discretionary{/}{%
}{/}S0377\discretionary{%
}{-}{-}0427\discretionary{%
}{(}{(}01\discretionary{)}{%
}{)}00485\discretionary{%
}{-}{-}X}}}


\bibitem{mayer2021visualization}
E.~Mayer, J.~McCullough, J.~Günther, S.~Cielo, and P.~Coveney.
\newblock Visualization of human-scale blood flow simulation using intel® ospray studio on supermuc-ng.
\newblock In {\em Proceedings of SC21: The International Conference for High Performance Computing, Networking, Storage, and Analysis}, 2021.

\bibitem{MoenneLoccoz2024}
N.~Moenne-Loccoz, A.~Mirzaei, O.~Perel, R.~de~Lutio, J.~Martinez~Esturo, G.~State, S.~Fidler, N.~Sharp, and Z.~Gojcic.
\newblock 3d gaussian ray tracing: Fast tracing of particle scenes.
\newblock {\em ACM Trans. Graph.}, 43(6),  article no. 232,  19 pages, Nov. 2024. \href{https://doi.org/10.1145/3687934}
{doi: {{%
10\hspace{.1pt}\discretionary{.}{%
}{.}\hspace{.4pt}1145\discretionary{/}{%
}{/}3687934}}}


\bibitem{Museth2014HDDA}
K.~Museth.
\newblock Hierarchical digital differential analyzer for efficient ray-marching in openvdb.
\newblock In {\em Proceedings of ACM SIGGRAPH Talks}, 2014.
\newblock Talk presented at SIGGRAPH 2014.

\bibitem{DBLP:conf/vmv/NiedermayrNPEW24}
S.~Niedermayr, C.~Neuhauser, K.~Petkov, K.~Engel, and R.~Westermann.
\newblock Application of 3d gaussian splatting for cinematic anatomy on consumer class devices.
\newblock In L.~Linsen and J.~Thies, eds., {\em 29th International Symposium on Vision, Modeling, and Visualization, {VMV} 2024, Munich, Germany, September 10-13, 2024}. Eurographics Association, 2024. \href{https://doi.org/10.2312/VMV.20241195}
{doi: {{%
10\hspace{.1pt}\discretionary{.}{%
}{.}\hspace{.4pt}2312\discretionary{/}{%
}{/}VMV\hspace{.1pt}\discretionary{.}{%
}{.}\hspace{.4pt}20241195}}}


\bibitem{NVIDIA2022DarkSky1M}
{NVIDIA}.
\newblock Dark sky 1m particles dataset.
\newblock \url{https://github.com/nvpro-samples/optix_advanced_samples/blob/master/src/data/darksky_1M.xyz}, 2022.

\bibitem{NvidiaOmniverse}
{NVIDIA Corporation}.
\newblock {\em {NVIDIA Omniverse}: A platform for collaborative 3D workflows}.
\newblock NVIDIA Corporation, 2025.

\bibitem{NVIDIAOptiX8Documentation}
{NVIDIA Corporation}.
\newblock Optix 8 ray tracing documentation.
\newblock \url{https://raytracing-docs.nvidia.com/optix8/index.html}, 2025.

\bibitem{OpenVDB}
{OpenVDB Contributors}.
\newblock Openvdb – a sparse volume data structure.
\newblock \url{https://www.openvdb.org/}, 2025.

\bibitem{Ravindran2017Houdini}
V.~Ravindran.
\newblock Visualizing astronomical data in houdini.
\newblock Master's thesis, Bournemouth University, UK, 2017.

\bibitem{DirectVolumeRendering2007}
{SCI‑VIS Course Team, University of Stuttgart}.
\newblock Direct volume rendering.
\newblock \url{https://cgl.ethz.ch/teaching/former/scivis_07/Notes/stuff/StuttgartCourse/VIS-Modules-06-Direct_Volume_Rendering.pdf}, 2007.

\bibitem{seo2024flod}
Y.~Seo, Y.~S. Choi, H.~S. Son, and Y.~Uh.
\newblock Flod: Integrating flexible level of detail into 3d gaussian splatting for customizable rendering.
\newblock {\em arXiv preprint arXiv:2408.12894}, 2024.

\bibitem{Sharma2025GaussianParticleApproximation}
I.~Sharma and D.~Schmalstieg.
\newblock 3d gaussian particle approximation of vdb datasets: A study for scientific visualization, 2025.
\newblock Submitted on 7 Apr 2025 (v1); Revised on 16 Apr 2025 (v2). \href{https://doi.org/10.48550/arXiv.2504.04857}
{doi: {{%
10\hspace{.1pt}\discretionary{.}{%
}{.}\hspace{.4pt}48550\discretionary{/}{%
}{/}arXiv\hspace{.1pt}\discretionary{.}{%
}{.}\hspace{.4pt}2504\hspace{.1pt}\discretionary{.}{%
}{.}\hspace{.4pt}04857}}}


\bibitem{talegaonkar2025vol3dgs}
C.~Talegaonkar, Y.~Belhe, R.~Ramamoorthi, and N.~Antipa.
\newblock Volumetrically consistent 3d gaussian rasterization.
\newblock In {\em Proceedings of the IEEE/CVF Conference on Computer Vision and Pattern Recognition (CVPR)}, pp. 10953--10963, June 2025.

\bibitem{vizzo2022vdbfusion}
I.~Vizzo, T.~Guadagnino, J.~Behley, and C.~Stachniss.
\newblock Vdbfusion: Flexible and efficient tsdf integration of range sensor data.
\newblock {\em Sensors}, 22(3):1296, 2022. \href{https://doi.org/10.3390/s22031296}
{doi: {{%
10\hspace{.1pt}\discretionary{.}{%
}{.}\hspace{.4pt}3390\discretionary{/}{%
}{/}s22031296}}}


\bibitem{Vucini2009PhD}
I.~Vučinić.
\newblock {\em Gaussian-Based Volumetric Reconstruction and Rendering Techniques}.
\newblock PhD thesis, Technische Universität Wien, 2009.

\bibitem{walker2022nanomap}
V.~Walker, F.~Vanegas, and F.~Gonzalez.
\newblock Nanomap: A gpu-accelerated openvdb-based mapping and simulation package for robotic agents.
\newblock {\em Remote Sensing}, 14(21):5463, 2022. \href{https://doi.org/10.3390/rs14215463}
{doi: {{%
10\hspace{.1pt}\discretionary{.}{%
}{.}\hspace{.4pt}3390\discretionary{/}{%
}{/}rs14215463}}}


\bibitem{ytini_amr}
{ytini}.
\newblock Amr data tutorial: The code, n.d.

\bibitem{inproceedings}
K.~Zhou, Q.~Hou, M.~Gong, J.~Snyder, B.~Guo, and H.-Y. Shum.
\newblock Fogshop: Real-time design and rendering of inhomogeneous, single-scattering media.
\newblock pp. 116 -- 125, 10 2007. \href{https://doi.org/10.1109/PG.2007.48}
{doi: {{%
10\hspace{.1pt}\discretionary{.}{%
}{.}\hspace{.4pt}1109\discretionary{/}{%
}{/}PG\hspace{.1pt}\discretionary{.}{%
}{.}\hspace{.4pt}2007\hspace{.1pt}\discretionary{.}{%
}{.}\hspace{.4pt}48}}}


\end{thebibliography}

\section*{Appendix A: Ray-Gaussian Intersection via Mahalanobis Distance}

To determine whether a ray intersects the ellipsoidal support of a 3D anisotropic Gaussian, we solve for the parameter \( t \) in the ray equation \( r(t) = v + t\hat{r} \) such that the Gaussian’s \textit{unnormalized} density remains above a fixed opacity threshold \( \kappa \in (0, 1) \). The unnormalized Gaussian density is defined as:
\[
\rho(x) = \exp\left( - (x - \mu)^\top \Sigma^{-1} (x - \mu) \right)
\]

The Gaussian’s support is bounded by the isosurface where this density falls to \( \kappa \), yielding the Mahalanobis distance condition:
\[
(x - \mu)^\top \Sigma^{-1} (x - \mu) \leq -\log \kappa
\]

Substituting the ray \( r(t) = v + t\hat{r} \) into the inequality gives:
\[
(v + t\hat{r} - \mu)^\top \Sigma^{-1} (v + t\hat{r} - \mu) \leq -\log \kappa
\]

Let \( \delta = v - \mu \). Expanding this yields a quadratic form in \( t \):
\[
Ct^2 + Bt + A + \log \kappa \leq 0
\]
where the coefficients are defined as:
\begin{align*}
	C &= \hat{r}^\top \Sigma^{-1} \hat{r} \\
	B &= 2\hat{r}^\top \Sigma^{-1} \delta \\
	A &= \delta^\top \Sigma^{-1} \delta
\end{align*}
Where, \( \delta = v - \mu \) denote the vector from the Gaussian center to the ray origin.

We solve the resulting quadratic equation:
\[
Ct^2 + Bt + A + \log \kappa = 0
\]

The real roots \( t_0 \) and \( t_1 \) define the entry and exit points of the ray through the Gaussian’s ellipsoidal support. If no real solutions exist, the ray does not intersect the Gaussian. If \( t_1 < 0 \), the intersection lies entirely behind the ray origin and can be ignored. The segment \( [t_0, t_1] \), if valid, is used to evaluate the line integral of the Gaussian density along the ray, as described in Equation~\eqref{eq:ray_gaussian_integral}.

\section*{Appendix B: Ray-Gaussian Line Integral}

The line integral of a scalar-valued Gaussian along a ray in 3D space is adopted from \cite{inproceedings}. This is implemented on GPU using CUDA’s intrinsic \texttt{erff()} function, which is hardware-accelerated and defined in \texttt{math\_functions.h}.

This allows us to perform efficient, closed-form evaluation of 1D Gaussian integrals along rays in the volume renderer.
Let the Gaussian be defined as:
\[
G(x) = c \exp\left(-a^2 \|x - b\|^2\right)
\]
where \( c \) is the peak amplitude, \( a \) controls the spatial spread (i.e., related to standard deviation as \( a = \frac{1}{\sqrt{2}\sigma} \)), and \( b \in \mathrm{R}^3\) is the Gaussian center.

We wish to evaluate the integral of \( G \) along the ray \( r(t) = v + t\hat{r} \) for \( t \in [0, d_r] \), where \( v \) is the ray origin, \( \hat{r} \) is the unit ray direction, and \( d_r \) is the ray segment length. Let:
\[
b' = b - v, \quad \hat{b}' = \frac{b'}{\|b'\|}
\]
and define the angle \( \xi \) between the ray direction and the vector to the Gaussian center:
\[
\cos\xi = \hat{r} \cdot \hat{b}', \quad \sin\xi = \|\hat{r} \times \hat{b}'\|
\]

Then the integral becomes:
\[
y = \int_0^{d_r} G(r(t)) \, dt = \int_0^{d_r} c \exp\left(-a^2 \|t\hat{r} - b'\|^2\right) dt
\]

This simplifies to:
\[
y = c \exp\left(-a^2 \|b'\|^2 \sin^2 \xi \right) \int_0^{d_r} \exp\left(-a^2 (t - \|b'\| \cos\xi)^2\right) dt
\]

Letting \( \mu = \|b'\| \cos\xi \), we get:
\[
y = c \exp\left(-a^2 \|b'\|^2 \sin^2 \xi \right) \int_0^{d_r} \exp\left(-a^2 (t - \mu)^2\right) dt
\]

This integral is the cumulative of a shifted Gaussian and can be evaluated using the error function:
\[
y = c \exp\left(-a^2 \|b'\|^2 \sin^2 \xi \right) \cdot \frac{\sqrt{\pi}}{2a} \left[ \operatorname{erf}(a(d_r - \mu)) - \operatorname{erf}(-a\mu) \right]
\]

\end{document}